\documentclass[journal]{IEEEtran}
\usepackage{cite}
\usepackage{algorithm}
\usepackage{algpseudocode}
\usepackage{array}
\usepackage{textcomp}
\usepackage{stfloats}
\usepackage{url}
\usepackage{verbatim}
\usepackage{graphicx}

\usepackage{amsmath,amssymb,amsfonts}
\usepackage{svg}
\usepackage{tikz}
\usepackage{pgfplots}
\pgfplotsset{compat=newest}
\pgfplotsset{plot coordinates/math parser=false}
\usepackage{diagbox}
\usepackage{array}
\usepackage{multirow}
\usepackage{import}
\usepackage{xcolor}
\def\BibTeX{{\rm B\kern-.05em{\sc i\kern-.025em b}\kern-.08em
		T\kern-.1667em\lower.7ex\hbox{E}\kern-.125emX}}
\begin{document}

\title{Automotive Radar Online Channel Imbalance Estimation via NLMS
}

\author{ \IEEEauthorblockN{Esmaeil Kavousi Ghafi\IEEEauthorrefmark{1}, Oliver Lang\IEEEauthorrefmark{1}, Matthias Wagner\IEEEauthorrefmark{1}, Alexander Melzer\IEEEauthorrefmark{2}, Mario Huemer\IEEEauthorrefmark{1} }\\
	\IEEEauthorblockA{\IEEEauthorrefmark{1}
		Institute of Signal Processing, Johannes Kepler University Linz, Austria}\\
	\IEEEauthorblockA{\IEEEauthorrefmark{2}
		Infineon Technologies Linz GmbH \& Co. KG, Austria\\
		Email: Esmaeil.Kavousi\_Ghafi@jku.at}
}

\maketitle

\begin{abstract}

Automotive radars are one of the essential enablers of advanced driver assistance systems (ADASs). Continuous monitoring of the functional safety and reliability of automotive radars is a crucial requirement to prevent accidents and increase road safety. One of the most critical aspects to monitor in this context is radar channel imbalances, as they are a key parameter regarding the reliability of the radar. These imbalances may originate from several parameter variations or hardware fatigues, e.g., a solder ball break (SBB), and may affect some radar processing steps, such as the angle of arrival estimation.
In this work, a novel method for online estimation of automotive radar channel imbalances is proposed. The proposed method exploits a normalized least mean squares (NLMS) algorithm as a block in the processing chain of the radar to estimate the channel imbalances. The input of this block is the detected targets in the range-Doppler map of the radar on the road without any prior knowledge on the angular parameters of the targets. This property in combination with low computational complexity of the NLMS, makes the proposed method suitable for online channel imbalance estimation, in parallel to the normal operation of the radar. 
Furthermore, it features reduced dependency on specific targets of interest and faster update rates of the channel imbalance estimation compared to the majority of state-of-the-art methods. This improvement is achieved by allowing for multiple targets in the angular spectrum, whereas most other methods are restricted to only single targets in the angular spectrum. The performance of the proposed method is validated using various simulation scenarios and is supported by measurement results.
\end{abstract}
\section{Introduction}

Equipping vehicles on the road with advanced driver assistance systems (ADASs), e.g., adaptive cruise control (ACC), autonomous emergency braking (AEB), and lane keep assistant (LKA), has significantly increased safety on the roads \cite{Gerstmair_safe,Gerstmair_Miniaturized}. ADASs exploit diverse types of sensors, such as cameras, lidars, and automotive radars \cite{b4}, to perceive the environment around the vehicle. However, automotive radars are considerably more robust than other sensors in harsh weather and lighting conditions. This is a key advantage of automotive radars, especially for long-range detection \cite{b5}. In addition, the low-cost and highly integrated implementation of mm-wave radars in monolithic microwave integrated circuits (MMICs) distinguishes them further from other sensors in terms of size and price \cite{b6}. These characteristics have made radars an essential component in ADASs. 

To mitigate the risk of hazardous traffic situations caused by random or systematic failures of software or hardware components in ADASs, it is required that all safety-critical components comply with stringent functional safety (FuSa) standards according to the ISO 26262 standard \cite{iso26262}. This standard encompasses a comprehensive range of FuSa aspects from FuSa management and product development to safety analyses of the electrical and electronic systems installed in road vehicles \cite{iso_overview}. 

For an automotive radar, as a part of ADASs, monitoring of the safety relevant parameters and nonidealities, as well as detection of faults is indispensable to achieve FuSa requirements. In this regard, it is crucial that the fault detection time is within the fault tolerant time interval (FTTI), ensuring that the system can respond promptly to faults and maintain a safe state \cite{bb_model}. Typically, FTTI for ADAS applications is considered to be up to few hundred milliseconds \cite{gulati2017ensuring, patent2020failure}. Examples of such online monitoring mechanisms in automotive radar systems are presented in \cite{Gerstmair_safe, bb_model,monitoring_wagner,monitoring_Melzer,Monitoring_Jeannin}.

In automotive radars that use multiple antennas on the transmit (Tx) and/or receive (Rx) side to measure the angular position of the objects, various nonidealities may distort the angular spectrum of the radar, thus, impairing its detection and tracking performance \cite{Guerzoni_movement}. These nonidealities may include imperfections in hardware (e.g., process variations), antenna mutual coupling, antenna array tolerances, temperature and voltage variations, device aging, and hardware damages such as a solder ball break (SBB) \cite{Effects_Schmid,calib_Grove,Impact_Geiss,bb_EKG}. They affect the angular spectrum through channel gain and phase differences, which, in this paper, are referred to as channel imbalances. 

Some of these nonidealities, e.g., imperfections in hardware, mutual coupling and array tolerances, are typically calibrated once due to their static nature. This so called end-of-line (EoL) calibration \cite{Requirements_Murad} is typically performed in a highly controlled environment, for instance in an anechoic chamber. Numerous works in the literature have been carried out to improve the calibration performance and also to reduce the requirements on the measurement setup and environment \cite{Guerzoni_movement,Geiss_calib, Durr_calib,Schmid_calib, SVDcalib}.

Other nonidealities, e.g., temperature and voltage variations, device aging and hardware damages, require online monitoring during the radar's operation on the road. This is due to their dynamic and random nature, which cannot be addressed through the EoL calibration. 
The automotive radar's installation behind the radome-emblem \cite{radome_calib} or the bumper \cite{bumper_calib,SelfCalibration2016} may also cause channel imbalances, especially since the position or characteristics of these covers may change over time.
Online mitigation approaches for these effects are proposed in \cite{radome_calib, bumper_calib,  af2011,SelfCalibration2016, AutoCalibration2021, af,CalibEKG, contrast_calib, ssfbcalib2024}. 

One of the first online channel imbalance estimation methods in the context of automotive radar was proposed in \cite{af2011}. Therein, the proposed method is based on a least squares (LS) method to estimate the complex gain imbalances across a linear virtual array (VA). For that, targets of interest defined as single targets in a range-Doppler (R-D) bin that have sufficiently high signal-to-noise ratio (SNR) are used. In this work, these targets are simply referred to as single targets. In \cite{SelfCalibration2016}, another method for online calibration of a three-dimensional beamforming automotive radar installed behind a bumper was proposed that also relies on single targets. The authors of \cite{AutoCalibration2021} propose a method based on simultaneous localization and mapping (SLAM). They exploit an extended Kalman filter (EKF) and the control inputs of the vehicle to address the joint SLAM and channel calibration problem. This method is not only restricted to single targets but also to stationary ones. A practical extension of \cite{af2011} was proposed in \cite{af} to estimate the channel imbalances. This method exploits a technique based on the stochastic gradient descent principle that aggregates the imbalances estimated from detected single targets, and thereby improves the estimation accuracy. The limitation of these methods is that they are restricted to single targets with sufficiently high SNR. Since they are scenario dependent as such, they may not be capable of responding within an FTTI in all scenarios. This can result in a hazardous situation, particularly for tasks that require fast response of the system, for instance in case of a hardware damage \cite{CalibEKG}.

A method that does not rely on specific targets of interest was presented in \cite{contrast_calib} for estimating the phase imbalances of the channels. This method is based on the optimization of several contrast metrics of the R-D map, e.g., variance and entropy. However, as mentioned in \cite{contrast_calib_base}, it cannot be claimed that for any particular R-D map or a set of R-D maps, maximizing the contrast provides the correct calibration. Also, the method may not be able to estimate the imbalances in certain cases, for instance, when the imbalances cause ghost targets. Furthermore, each iteration of the exploited optimization algorithm requires the computation of an R-D map, typically using a two-dimensional fast Fourier transform (2D FFT), resulting in high computational complexity.

A subspace-based method for estimating the channel imbalances and mutual coupling was presented in \cite{ssfbcalib2024}. The core idea of this method is that, in the presence of channel imbalances and mutual coupling, the signal subspace leaks into the noise subspace of the autocorrelation matrix of the received data. Accordingly, authors apply a nonparametric model order selection approach to minimize the leakage and estimate the channel imbalances and mutual coupling coefficients. An advantage of this method is that it is not restricted to single targets, and although it estimates the number of targets in each R-D bin to find the signal subspace dimensions, it does not require estimating target parameters. However, this method relies on a numerical optimization algorithm, for which in each iteration multiple eigenvalue decompositions are performed. As a result, the computational complexity of the algorithm is high, making it impractical for real-time applications, such as online calibration of channel imbalances and SBB detection. 

Another online method that is not restricted to single targets was proposed in \cite{CalibEKG}. This method employs a LS-based approach and exploits the periodic behavior of imbalances on the VA to estimate them. The drawback of this method is that it is restricted to Rx channel imbalance estimation.

In this paper, we propose a novel online channel imbalance estimation method for uniform linear array (ULA) automotive radars. This method leverages the well-known signal processing approach of normalized least mean squares (NLMS) adaptive filtering. 
We model the complex channel gains with a set of single-tap filters, each filter modeling one of the channel gains with a complex weight. A corresponding set of single-tap NLMS filters, equal in number to the channels, is applied to estimate the filters weights. 
The samples at each detected peak from the R-D maps and across the VA are considered as the inputs to the algorithm. In this paper, we refer to these samples in the vector form as a signal vector. These signal vectors represent the output of the single-tap filters, which are distorted by the imbalances. As the NLMS filters require the input of the single-tap filters as input, a reconstruction block is required to estimate that from the signal vectors. To this aim, after calibrating the signal vector using the most recent estimate of the imbalances, a parameter estimation method is employed to estimate the parameters of each target, namely angular frequency and complex amplitude. These parameters are then used to reconstruct the undistorted signal vector at the input of the NLMS filters.
The estimated parameters in the reconstruction block can be considered as the outputs of a typical DoA block. And, the estimated complex gains from the NLMS block are used for channel calibration, and to determine the Rx and Tx gain and phase imbalances (GPIs) for a hardware damage detection, e.g., a SBB detection.

The key advantages of the proposed method in this work can be summarized as follows:
\begin{itemize}
	\item The proposed method does not require any prior knowledge of the targets in the environment, as it estimates the targets' parameters. Therefore the method can be considered as a blind imbalance estimation method \cite{ssfbcalib2024}.
	\item In contrast to the major part of the literature, the proposed method is not restricted to certain targets of interest such as single targets. That is, a great number of signal vectors containing multiple targets, which are discarded by those methods, are being processed in our method. This enables the proposed method to continuously perform channel imbalance estimation even in scenarios where single targets are rare or missing. 
	\item Being unrestricted to single targets, and as a result, having more input data, the proposed method can perform faster in estimating the channel imbalances, compared to using only single targets.
	\item The computational complexity of the method is low, since its main processing block is the NLMS. It should be noted that the computation complexity of the signal reconstruction block can be excluded from that of the imbalance estimation, since it is already required for the DoA estimation block in the normal processing chain of an automotive radar. Therefore, one can conclude that the proposed method is highly suitable for online channel imbalance estimation of the radar, in parallel to its normal operation.
\end{itemize}

The remainder of the paper is organized as follows. In Section \ref{model}, the signal and system model, the problem formulation, and preprocessing steps are presented. Section \ref{method} introduces the proposed channel imbalance estimation method. Important aspects of the proposed method are discussed in details in Section \ref{considerations}. In Section \ref{results}, the results of several simulations are presented to validate the performance of the proposed method, followed by Section \ref{measurement_results} in which the method is applied to measurement data. Finally, Section \ref{conclusion} concludes the work.

 \section{Signal and system model}\label{model}
In the first part of this section, we introduce the automotive radar signal processing that we consider in this paper. In the second part, the channel imbalances with their sources and effects on the performance of the radar are described. The DoA bias problem, as one of the negative impacts of the channel imbalances, is reviewed in the third part.

\subsection{Automotive Radar Signal Processing}
\begin{figure*}[t]
	\centering
	\includegraphics{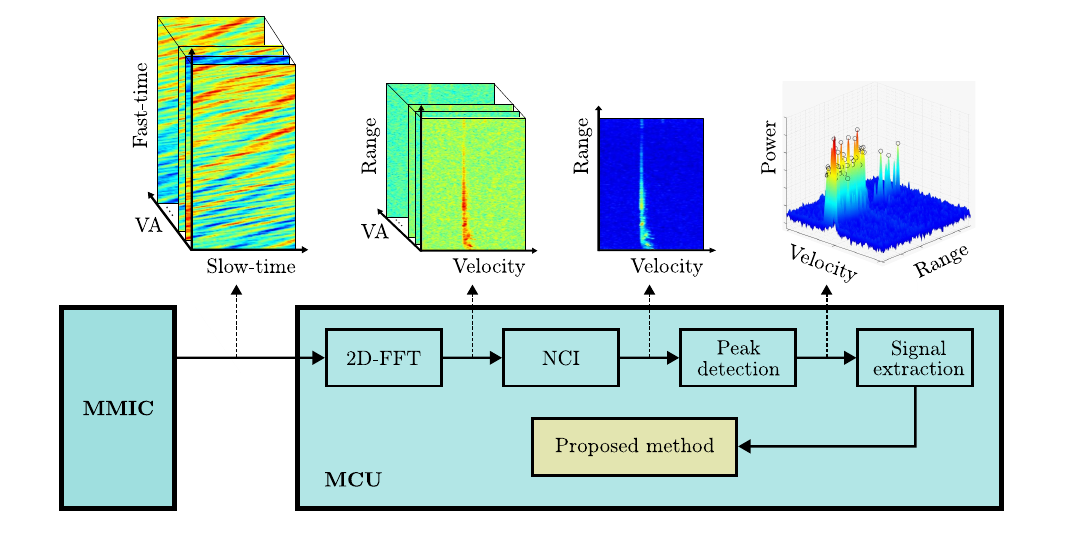}
	\caption{ Signal processing block diagram of an FMCW radar. The proposed method is a block on top of the normal processing flow of the radar.}
	\label{block_processing}
\end{figure*}

 In this paper, we consider a conventional frequency modulated continuous wave (FMCW) radar that incorporates only one MMIC. However, the proposed method can easily be extended to other types of automotive radars and with more MMICs, as in cascaded systems. 
 In an automotive FMCW radar, signals collected by the Rx antennas are mixed with the instantaneous Tx signal and subsequently filtered by low-pass filters. Afterwards, the resulting intermediate frequency (IF) signals that are sampled with analog-to-digital converters (ADCs) are delivered to a processor, e.g., a microcontroller unit (MCU), for digital signal processing. The digital signals at the MCU input are conventionally represented in a three-dimensional tensor known as a radar cube, characterized by the dimensions fast-time, slow-time, and VA elements. In a conventional FMCW radar signal processing pipeline, as shown in Fig.~\ref{block_processing}, first, a 2D-FFT is applied to the fast- and slow-time dimensions of the radar cube, generating a processed cube containing all R-D maps. Then, a non-coherent integration (NCI) of R-D maps across the VA is performed to create a single R-D map corresponding to the current processing cube. This integrated R-D map is then processed by a peak detection algorithm, such as a hard thresholding method or a constant false alarm rate (CFAR) algorithm \cite{CFAR}, to detect the targets and estimate their corresponding R-D indices. 
 An exemplary ideal signal vector along the VA dimension of the processed cube, specified by these indices, can be represented as $\mathbf{s} = [s[1], s[2],...,s[K]]^\text{T} $ with
\begin{equation}\label{s}
	{s}[k] = \sum_{q=1}^{Q} \alpha_q e^{\text{j} (2 \pi f_q^\theta k )} ,
\end{equation}
where $K = K_\text{t} K_\text{r}$ is the size of the VA, with $K_\text{t}$ and $K_\text{r}$ as the Tx and Rx array sizes, respectively. $Q$ is the number of targets in the corresponding R-D bin. 
$\alpha_q$ represents the complex amplitude, including the constant gain and phase, and $f_q^\theta$ denotes the angular frequency of the $q$th target. Also, j is the imaginary unit. These signal vectors are extracted from the processed cube and input to the proposed method. It should be noted that in an FMCW radar that employs time division multiplexing (TDM), the velocity-induced phase shifts \cite{phase_comp,TDM_Jeannin} have to be compensated before the proposed method is applied.

\subsection{Imbalance}\label{subsec:Imbalance}
 
The model in \eqref{s} is an ideal case, in which the Tx and Rx channels, including all their components, are not affecting the signals. In practice, each Tx and Rx channel scales the signal passing through by a complex gain factor and additive noise. Therefore, a more realistic model of the VA signal vector can be introduced as 

\begin{equation}\label{x}
	\mathbf{x} = \boldsymbol{\Psi} \mathbf{s} + \mathbf{n},
\end{equation}
where $\mathbf{x} = [x[1], x[2],...,x[K]]^\text{T}$, and $\mathbf{n} = [n[1], n[2],...,n[K]]^\text{T}$ is a complex zero-mean noise vector. $\boldsymbol{\Psi} = \text{diag}(\boldsymbol{\psi} )$, where $\boldsymbol{\psi}  = [\psi [1], \psi [2],...,\psi [K]]^\text{T}$ is the complex gain offset vector on the VA, and $\text{diag}(.)$ returns a square diagonal matrix with the elements of the input vector on the main diagonal.

Ideally, the complex gain offset values are equal across the channels. The more these values deviate from each other, the more the actual signal vector received from the environment deteriorates. Therefore, their relative values compared to a reference are more of interest rather than their absolute values. These relative values, referred to as channel imbalances, are denoted by $\boldsymbol{\xi}=[\xi[1], \xi[2],...,\xi[K]]^\text{T}$, where

\begin{subequations}\label{imbalance}
	\begin{align}
	 \boldsymbol{\xi} &= \frac{\boldsymbol{\psi}}{{\psi} [1]} \label{imbalance1}, \\ 
	                    &= \boldsymbol{\xi}_{\text{t}} \otimes \boldsymbol{\xi}_{\text{r}},\\
	 \boldsymbol{\xi}_{\text{t}} &=  [\xi_\text{t}[1], \xi_\text{t}[2],...,\xi_\text{t}[K_\text{t}]]^\text{T},\\
	 \boldsymbol{\xi}_\text{r}   & = [\xi_\text{r}[1], \xi_\text{r}[2],...,\xi_\text{r}[K_\text{r}]]^\text{T},
	 \end{align}
\end{subequations}
where $\otimes$ denotes the Kronecker product. In \eqref{imbalance}, $\boldsymbol{\xi}_{\text{t}}$ and $\boldsymbol{\xi}_\text{r}$ are Tx and Rx imbalances with $\xi [1] = \xi_\text{t}[1] = \xi_\text{r}[1] = 1$. 

The VA, Tx and Rx GPIs can also be introduced through $\gamma[k]$ and $\varphi [k]$ with $\xi[k] =(1 + \gamma[k])e^{\text{j}\varphi [k]}$, $\gamma_\text{t}[k_\text{t}]$ and $\varphi_\text{t} [k_\text{t}]$ with $\xi_\text{t}[k_\text{t}] = (1+\gamma_\text{t}[k_\text{t}])e^{\text{j}\varphi_\text{t} [k_\text{t}]}$, and $\gamma_\text{r}[k_\text{r}]$ and $\varphi_\text{r} [k_\text{r}]$ with $\xi_\text{r}[k_\text{r}] =(1+\gamma_\text{r}[k_\text{r}])e^{\text{j}\varphi_\text{r} [k_\text{r}]}$, respectively.

These imbalances may originate from a variety of sources including: 
\begin{itemize}
	\item Imperfections in hardware: during the manufacturing process, the same components in different channels might be produced with some variation. For example, differences may occur in the physical length of antennas or in the characteristics of filters.
	\item Antenna array tolerances: when the physical location and inter-element distance of the antennas has some tolerances relative to the array model. This issue introduces DoA-dependent imbalances to channels.  
	\item Temperature and voltage variations: all components on the MMIC are susceptible to temperature and voltage fluctuations. As different channels might respond to these fluctuations differently, channel imbalances may occur. In cascaded radar systems, this effect is more significant, as the response of channels to the fluctuations across different MMICs may vary more significantly compared to those within each individual MMIC.
	\item Device aging: similar to any other electronic devices, components and parameters of the radar are vulnerable to aging, and vary during the lifetime of the sensor. These variations also may cause channel imbalances.
	\item Hardware damages: these damages, for instance, can be a failure in one of the components or a SBB.
	\item Radar covers: automotive radars are conventionally installed behind the bumper or the radome-emblem of the vehicle. Non-homogeneous shape, painting or coating of this covers, disturbs the normal signal paths to the antennas causing channel imbalances. 
\end{itemize}

In this paper, we assume static channel imbalances originating from imperfections in hardware, mutual coupling and array tolerances are calibrated in the EoL calibration process. Here, we consider solely the dynamic channel imbalances. These imbalances are not DoA dependent, and their values are bounded within certain limits. Here, we also consider a SBB detection as an example for the hardware damage detection problem. We model a SBB as a significant phase imbalance, e.g., $30^o$, on one of the channels \cite{bb_model,bb_EKG}.

These imbalances deteriorate the detection of the targets in the angular domain, mainly through sidelobe level (SLL) increment and DoA estimation errors. To investigate these problems further, one can separate the channel phase imbalances $\varphi[k]$ into two terms; a linear phase progression with respect to $k$, and a residual term $\varphi_\Delta[k]$ as
 
 \begin{equation}\label{df}
 		\varphi[k] = 2\pi f_\Delta k + \varphi_\Delta[k],
 \end{equation}
 where $2\pi f_\Delta$ is the slope of the linear phase. Combining \eqref{s}, \eqref{x}, and \eqref{df}, and applying mathematical simplifications yields
\begin{equation}\label{doa_e}
	\begin{aligned}
		x[k]  		&=   \psi[1] (1 + \gamma[k])e^{\text{j} \varphi_\Delta[k]} \sum_{q=1}^{Q} \alpha_q e^{\text{j} (2 \pi (f_q^\theta + f_\Delta) k )} \\
		& \hspace{5cm}+ n[k],
	\end{aligned}
\end{equation}

As can be seen from \eqref{doa_e}, the linear term of the phase imbalances shifts the angular frequency of the targets by $f_\Delta$. Thereby, the channel phase imbalances lead to a DoA bias in case $\varphi [k]$ has a linear trend over $k$. 

\begin{figure}[t]
	\centering
	\input{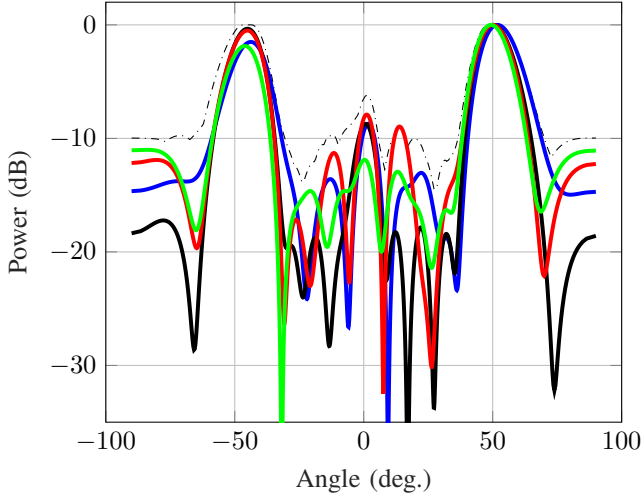}
	\caption{ The angular spectra corresponding to a scenario with three targets and random channel imbalances. The angular spectra represent; black: the imbalance-free angular spectrum, red: ghost targets, green: target suppression, and blue: DoA bias. The dashed-line spectrum corresponds to the maximum value of all MCSs.}
	\label{imbspec}
\end{figure}

In addition, it can be shown that the residual imbalance term $(1 + \gamma[k])e^{\text{j} \varphi_\Delta[k]}$ increases the SLL. This increment in SLL reduces the radar's dynamic range, potentially obscuring targets with low SNR. Moreover, the interactions between the imbalance vector and the original signal vector may create ghost targets, suppress targets in the spectrum, and degrade the resolution in case of two closely spaced targets. To illustrate these effects, a signal vector consisting of three targets at $-45^o$, $0^o$ and $50^o$ is considered. This signal is deteriorated by random GPIs uniformly distributed in the range of ${\varphi_\text{t}, \varphi_\text{r}} \in [-20^o \hspace{1mm} 20^o]$ and ${\gamma_\text{t}, \gamma_\text{r}} \in [-0.2 \hspace{1mm} 0.2]$. From one hundred Monte Carlo simulations (MCSs), the worst case scenarios for each prescribed effect are selected, and their corresponding spectra are plotted in Fig.~\ref{imbspec}. Further detailed analyses on the modeling and effects of channel imbalances can be found in \cite{Effects_Schmid, Grove_Impact, Eschbaumer_Impact, Impact_Geiss}.

Considering \eqref{doa_e}, since the proposed algorithm is a blind approach without any prior knowledge on the targets' frequencies $f_q^\theta$, DoA estimations are biased by an angle corresponding to $f_\Delta$. Therefore, the proposed method is not able to estimate the linear term of the phase imbalances, i.e., $2\pi f_\Delta k$.

\subsection{DoA Bias Estimation}

In \cite{bumper_calib}, a method to estimate the angle-dependent DoA bias for the radars installed behind a bumper was proposed. In addition, one can address the DoA bias estimation problem similar to the radar misalignment problem \cite{alignment_Kellner}. The alignment of the radar sensor mounted on the vehicle may deviate from the desired alignment. As a result, the scenery is rotated by this misalignment angle, introducing a systematic DoA bias equal to this misalignment angle. This problem has been thoroughly investigated in the literature \cite{alignment_Kellner, alignment_Ikram, alignment_Bao, alignment_Bobaru, alignment_radarnetwork, alignment_SensorX2Vehicle}. Their fundamental idea is to exploit the geometrical correlation between the DoA of the targets in the scenery (typically the stationary ones) and the ego vehicle's motion parameters, e.g., ego velocity. A straightforward illustration of this relation can be found in \cite{ego_motion}. Nevertheless, the main drawback of these approaches is that the measurements from each isolated radar are not sufficient for the estimations, and further information from other radars or modalities is required. For instance, vehicle's motion parameters, e.g., ego velocity, from the odometry sensors.

A straightforward method to estimate the DoA bias, considered as a superposition of the DoA biases resulting from channel imbalances and radar misalignment, is described as follows. This method is based on the aforementioned works in this subsection, and relies solely on measurements from a single radar. Assuming the ego vehicle, equipped with a front-looking radar, is moving in a straight line with ego velocity $v_s$, the relation between actual DoA of stationary targets $\theta^i$, where the superscript $i$ is the index of the targets, and ego velocity is as follows \cite{directionsandchallenges}

\begin{equation}\label{v}
	v_t^i = v_s \cos(\theta^i + \theta_b),
\end{equation}
where $v_t^i$ are the measured relative velocities of the stationary targets, and $\theta_b$ is the DoA bias. Considering \eqref{v}, a cosine function is first fitted to the multiple measurements as $(\theta^i + \theta_b,v_t^i)$. Then, $v_s$ is determined by computing the maximum value of the cosine function, and $\theta_b$ is determined by finding the angle at which this maximum occurs. Clearly, the accuracy of this approach depends highly on the validity of the assumptions and the accuracy of the applied curve fitting.

In the rest of the paper, for simplicity, we assume the linear term of the channel imbalances is zero, i.e., $f_\Delta =0$, and we focus solely on the estimation of the residual part of the channel imbalances, i.e., $\xi[k] = (1 + \gamma[k])e^{\text{j} \varphi_\Delta[k]}$, for all $k$.

\section{Proposed Method}\label{method}
The model in \eqref{x} consists of the actual signal representing the reflections from targets $\mathbf{s}$ as well as the offset vector $\boldsymbol{\psi}$. One can consider \eqref{x} as a model for $K$ single-tap filters with the inputs $s[k]$, output $x[k]$ and the filters’ coefficients $\psi[k]$, where $k=1, 2,...,K$. The outputs of the filters are then deteriorated further by additive noise $n[k]$. Therefore, estimating the offset vector and, consequently, the channel imbalances is equivalent to the estimation of the filters coefficients. 

\begin{figure}
	\centering
	\includegraphics{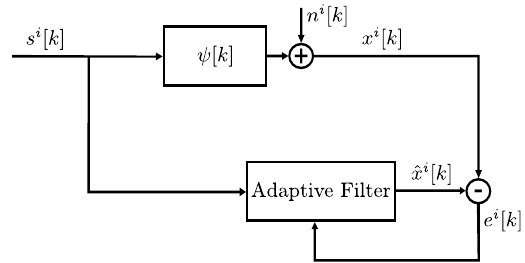}
	\caption{ Block diagram of system identification using an adaptive filter.}
	\label{block_af}
\end{figure}

\begin{figure}[b]
	\centering
	\includegraphics{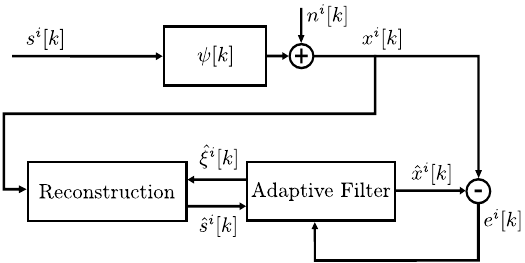}
	\caption{Block diagram of the proposed method.}
	\label{block_x_af}
\end{figure}

In order to estimate each filter's coefficient, a single-tap adaptive filter is utilized. Such an approach, as illustrated in Fig.~\ref{block_af}, is referred to as system identification in the signal processing literature \cite{diniz1997adaptive}. Typically, in a conventional adaptive filtering approach for system identification, the input $s[k]$ is known. However, in our case it is unknown and only its filtered version is available. To tackle this issue, we employ an additional processing block, denoted as reconstruction block, to estimate the input signal. Therefore, the proposed method consists of a cyclic approach with two steps in each iteration, as shown in Fig.~\ref{block_x_af}. 
First, a reconstruction block estimates $\hat{s}^i[k]$ values, with $i$ being the iteration index, using the radar model \eqref{s} and the current estimate of the channel imbalances $\hat{\xi}^i[k]$. Second, an adaptive filter block estimates $\hat{\xi}^i[k]$ values using the estimated $\hat{s}^i[k]$ values. Note that at each new iteration $i+1$, a new signal vector $\mathbf{x}^{i+1}$ is input to the algorithm. In the following subsections, we explain these blocks further. 

\subsection{Reconstruction}
In the reconstruction step, the portion of the signal vector $\mathbf{x}^i$ that corresponds to the reflections from the objects in the environment, i.e., $\mathbf{s}^i$, is estimated. It should be noted that, although the filters coefficients are estimated separately across the channels, the reconstruction block is the same for all estimations. 

The reconstruction of $\mathbf{s}^i$ is generally carried out in three steps of predistortion, parameter estimation, and integration as follows.

\subsubsection*{Predistortion}
A predistortion step is required to mitigate the effect of the channel imbalances on the signal vector as much as possible using a calibration vector corresponding to that iteration. The calibration vector is computed simply by the element-wise inversion of the most recent imbalance estimates as $\mathbf{c}^i = 1 \oslash \hat{\boldsymbol{\xi}}^i$, where $\oslash$ denotes the element-wise division. The initial calibration vector at $i=1$ is set to an all-ones vector. The resulting signal after predistortion can be written as follows
 \begin{equation}\label{predistortion}
		\mathbf{x}^i_\text{pd} = \mathbf{C}^i  \mathbf{x}^i,
\end{equation}
where $\mathbf{C}^i = \text{diag} (\mathbf{c}^i)$.

\subsubsection*{Parameter Estimation and Integration}
This step estimates the signal $\mathbf{s}^i$ from the noisy signal $\mathbf{x}^i_\text{pd}$ through estimating its parameters. These parameters based on the signal model \eqref{s}, include $Q$, as well as $\alpha_q$ and $f_q^\theta$ for all $q$, and their estimates are denoted as $\hat{Q}$, $\hat{\alpha}_{u}$ and $\hat{f}_{u}^\theta$ for $u = 1,2,...,\hat{Q}$. 
Possible methods to estimate these parameters include the FFT \cite{CalibEKG}, the CLEAN algorithm \cite{clean}, the RELAX algorithm \cite{relax}, the MUSIC algorithm \cite{music}, and the sparse cyclic coordinate descent (SCCD) algorithm \cite{Garcia2021}.

After estimation of the parameters, $\mathbf{s}^i$ is reconstructed as $\hat{\mathbf{s}}^i = [\hat{s}^i[1], \hat{s}^i[2], ... , \hat{s}^i[K]]^T$ with

\begin{equation}\label{s_hat}
	\hat{s}^i[k] = \sum_{u=1}^{\hat{Q}} \hat{\alpha}_{u} e^{\text{j} (2 \pi \hat{f}_{u}^\theta k )}.
\end{equation}

\begin{figure}[t]
	\centering
	\includegraphics{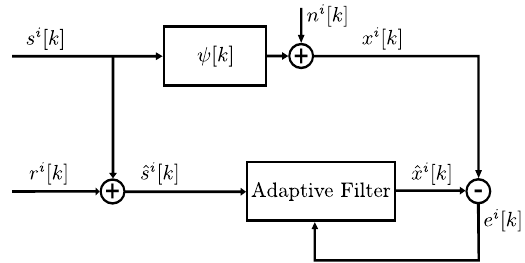}
	\caption{Adaptive filter with noisy input.}
	\label{inputnoise_bd}
\end{figure}

The error between the original signal $\mathbf{s}^i$ and the reconstructed one $\hat{\mathbf{s}}^i$, computed as
\begin{equation}\label{r}
	\mathbf{r}^i = \hat{\mathbf{s}}^i - \mathbf{s}^i,
\end{equation}
may generally degrade the estimation of the filters coefficients. The reconstruction error $\mathbf{r}^i$ can be considered as an additive noise at the input of the adaptive filters, as shown in Fig.~\ref{inputnoise_bd}. This noise at the input of the adaptive filters leads to a biased estimation of the channel imbalances. A detailed derivation of the bias vector is provided in Subsection \ref{convergence}.

Such problem, referred to as adaptive filtering with noisy input, has been addressed in multiple works such as \cite{noisyinput2005,noisyinput2013,noisyinput2019,noisyinput2023,noisyinput2024}. Solutions proposed in these works attempt to derive a compensation term to be added to the update equation of the adaptive filters to compensate for the bias in the final estimates. In these solutions, the input noise is typically assumed to be independent of the input signal. Also, the compensation terms are functions of the input noise variance $\sigma^2_{r^i[k]}$ estimated in a separate step. However, in our practical problem, this independence assumption is not valid and the input noise $\mathbf{r}^i$ is correlated with the input signal $\mathbf{s}^i$. Consequently, computation of the compensation term is challenging. Therefore, we do not compensate the bias in this paper. Instead, we employ parameter estimation algorithms that lead to acceptable reconstruction errors and thus add tolerable biases to the imbalance estimates.

Although FFT can be considered as the most straightforward algorithm for the parameter estimation, it leads to large reconstruction errors, and hence significant biases. The main reason is that sidelobes of each target peak in the angular spectrum interfere the mainlobe of the other peaks, resulting in erroneous estimations of the parameters. Therefore, in order to overcome this issue and decrease the reconstruction error, a more sophisticated method should be applied. In this paper, we utilize the CLEAN algorithm, which is an iterative algorithm based on the FFT. In each iteration, the dominant peak in the spectrum, is extracted and coherently removed from the signal before the FFT. The process continues until a predefined criterion is fulfilled, e.g., the power of the most recently detected peak is lower than a predefined threshold. In this manner, the target interference problem of the FFT is addressed effectively, and thereby the reconstruction error and resulting biases are reduced significantly. A pseudo-code of the CLEAN algorithm is provided in Algorithm \ref{clean_alg}. In this Algorithm, $\hat{\boldsymbol{\alpha}}$ and $\hat{\mathbf{f}}^\theta$ denote the vectors containing $\hat{Q}$ complex amplitudes and frequencies, respectively. $P$ is the threshold used for the termination criterion, $N$ is the FFT length, and $|\cdot|$ computes the absolute value. max($\cdot$) returns the maximum value of the input vector, and find($\cdot$) is a function to find the index at which its argument is satisfied. Squared brackets are used for indexing, and indices start from 1. The inter-element spacing of the VA is set to $d=\frac{\lambda}{2}$, where $\lambda$ is the wavelength of the radar signal. Also, $\mathbf{x}_{\text{r}} = [x_{\text{r}}[1], x_{\text{r}}[2],...,x_{\text{r}}[K]]^T$.

\begin{algorithm} 
	\caption{Parameter Estimation via the CLEAN algorithm}
	\label{clean_alg}
	\begin{algorithmic}[1] 
		\State \textbf{Input:} $\mathbf{x}^i_{\text{pd}}$, $N$, $P$ 
		\State \textbf{Output:} $\hat{Q}$, $\hat{\boldsymbol{\alpha}}$ and $\hat{\mathbf{f}}^\theta$
		\State $K \gets \text{length}(\mathbf{x}^i_{\text{pd}})$
		\State $\hat{\boldsymbol{\alpha}} \gets [\hspace{0.1cm}]$
		\State $\hat{\mathbf{f}}^\theta \gets [\hspace{0.1cm}]$
		\State $\hat{Q} \gets 1$
		
		\While{true} 
		\State $\mathbf{y} \gets \text{FFT}(\mathbf{x}^i_{\text{pd}},N)$ {\color{gray}\small \Comment{$N$-point FFT}}
		\State $ l \gets \text{find}(|\mathbf{y}| = \text{max}(|\mathbf{y}|))$
		\State $\hat{{\alpha}} \gets y[l]$
		\State $\hat{{f}}^\theta \gets -0.5 + (l-1)/N$
		\For{$k \gets 1$ to $K$}
		\State $x_{\text{r}}[k] \gets \hat{{\alpha}}e^{\text{j}\left(2\pi \hat{{f}}^\theta (k-1)\right)}$
		\EndFor
		\State $\mathbf{x}^i_{\text{pd}} \gets \mathbf{x}^i_{\text{pd}} - \mathbf{x}_{\text{r}}$
		\If{$\hat{Q} > 1$}
			\If{$\frac{| \hat{{\alpha}} |} { | \hat{\boldsymbol{\alpha}}[1]| } < P$} 
			\State \textbf{Break}
			\EndIf
		\EndIf
		\State $\hat{\boldsymbol{\alpha}}[\hat{Q}] \gets \hat{{\alpha}}$
		\State $\hat{\mathbf{f}}^\theta [\hat{Q}] \gets \hat{{f}}^\theta$
		\State $\hat{Q}  \gets \hat{Q} +1$
		\EndWhile 
	\end{algorithmic}
\end{algorithm}

\subsection{Adaptive Filter}
After the reconstruction of the input signal vector, multiple single-tap adaptive filters are employed to estimate the channel imbalances. In this paper we exploit a widely used NLMS \cite{sayed2003fundamentals} for this task. The cost function of the NLMS is the instantaneous squared error between the measurement $x^i[k]$ and the output of the adaptive filter $\hat{x}^i[k] = \hat{\psi}^i [k] \hat{s}^i[k] $ as
\begin{equation}\label{j}
			J^i[k]  = \frac{1}{2} (\hat{\psi}^i [k] \hat{s}^i[k] - x^i[k])(\hat{\psi}^i [k] \hat{s}^i[k] - x^i[k])^\ast,
\end{equation}
where $(\cdot)^\ast$ denotes the complex conjugate, and $\hat{\psi}^i [k]$ is the $k$th NLMS estimate in the estimation vector $\hat{\boldsymbol{\psi}}^i = [\hat{\psi}^i[1],\hat{\psi}^i[2],...,\hat{\psi}^i[K]]^T$. Incorporating the steepest descent optimization method yields the update equation of the NLMS as \cite{sayed2003fundamentals}
\begin{equation}\label{update_equation}
	\hat{\psi}^{i+1} [k]  = \hat{\psi}^i[k]  - \mu^i \nabla J^i[k],
\end{equation}
where $\nabla J^i[k]$ and $\mu^i$ are gradient and step size of the NLMS, and are computed as \cite{sayed2003fundamentals}
\begin{equation}\label{grad}
	\nabla J^i[k] = (\hat{s}^i [k])^\ast (\hat{\psi}^i[k]  \hat{s}^i[k] - x^i[k]),
\end{equation}
and 
\begin{equation}\label{mu}
	\mu^i = \frac{\mu_0}{(\hat{\mathbf{s}}^i)^H \hat{\mathbf{s}}^i},	
\end{equation}
where $(\cdot)^H$ denotes the Hermitian transpose, and $\mu_0$ denotes the normalized step size. It should be noted that $\mu^i$ is a scalar value, and the same for every update equation across channels. The reason that we chose this step size, rather than different step sizes for different channels, for example in the form of $\mu^i[k] = \frac{\mu_0[k]}{(\hat{s}^i[k])^\ast \hat{s}^i[k] + \epsilon [k]}$, is as follows. First, $K$ normalized step sizes $\mu_0[k]$ would have to be set. Second, since the term $(\hat{s}^i[k])^\ast \hat{s}^i[k]$ may get close to zero, $K$ regularization values $\epsilon [k]$ have to be set to guarantee the stability of the algorithm. Consider that in \eqref{mu}, $(\hat{\mathbf{s}}^i)^H \hat{\mathbf{s}}^i \gg 0$. Third, even though different step sizes across channels may result in faster convergence of the NLMS for some channels, the overall performance of the algorithm in terms of convergence and estimation error is not improved notably. Nevertheless, the choice of step size in \eqref{mu} relaxes the requirement for multiple hyperparameters of $\mu_0[k]$ and $\epsilon [k]$. Even if the same hyperparameters are selected for different channels, an additional hyperparamter of $\epsilon$ is still required compared to the step size in \eqref{mu}.

As the estimates of the adaptive filters are looped back to the reconstruction block, a time varying ambiguity $\lambda^i[k]$ may be observed such that  $\hat{x}^i[k] = (\lambda^i[k] \hat{\psi}^i [k]) (\hat{s}^i[k] \frac{1}{\lambda^i[k]}) $. Generally, this may distort the estimates in two ways: a monotonic attenuation of the estimates, and adding a time varying linear phase trend to the estimated vector's phase. To address this ambiguity, a normalization and a detrend function are proposed to equalize the attenuation and the phase trend in the feedback in each iteration as follows. A similar step is also applied in \cite{af}. 
\begin{subequations}\label{normalisation}
	\begin{align}
		\hat{\boldsymbol{\psi}}^{i+1}_n &= \frac{\hat{\boldsymbol{\psi}} ^{i+1}}{\hat{\psi} ^{i+1}[1]},\\ \label{normalisation_1}
		\hat{\boldsymbol{\varphi}}^{i+1} &=    \text{detrend}(\angle \hat{\boldsymbol{\psi}}^{i+1}_n ),
	\end{align}
\end{subequations}
where $\hat{\boldsymbol{\varphi}}^{i+1} = [\hat{\varphi}^{i+1}[1],\hat{\varphi}^{i+1}[2],...,\hat{\varphi}^{i+1}[K]]^T$, and $\angle$ extracts and unwraps the phase.
$\text{detrend}(\cdot)$ removes the linear trend in the input in the form of
\begin{equation}\label{line}
	\mathbf{z} = p_1 \mathbf{y} + p_2,
\end{equation}
where $\mathbf{y} = [1,2,...,K]^T$, and $p_1$ and $p_2$ are the slope and the intercept point of the line, respectively. One can estimate these parameters using a least squares (LS) estimation as 
\begin{equation}\label{ls}
	\hat{\mathbf{p}} = (\mathbf{Y}^T \mathbf{Y})^{-1} \mathbf{Y}^T (\angle \hat{\boldsymbol{\psi}}^{i+1}_n),
\end{equation}
where 
\begin{subequations}\label{ls_parameters}
	\begin{align}
		\mathbf{Y} &= 
		\begin{bmatrix}
			1 & 2 & ... & K \\
			1 & 1 & ... & 1
		\end{bmatrix} ^ T ,\\
		\hat{\mathbf{p}} &= [\hat{p_1} \hspace{0.2cm} \hat{p_2}]^T.
	\end{align}
\end{subequations}

Finally, the channel imbalance estimates are determined as
\begin{equation}
	\begin{aligned}
		\hat{\boldsymbol{\xi}}^{i+1}  &= (1+ \hat{\boldsymbol{\gamma}}^{i+1}) \odot e^{\text{j} \hat{\boldsymbol{\varphi}}^{i+1}}
	\end{aligned}
\end{equation}
where $\hat{\boldsymbol{\gamma}}^{i+1} = |\hat{\boldsymbol{\psi}}^{i+1}_n | - 1$, and $\odot$ denotes the Hadamard product.

In a typical radar processing, the estimated channel imbalances $\hat{\boldsymbol{\xi}}^{i+1} $ can be used for the calibration of the next input signal vector. However, this calibration step is continuously being carried out in the predistortion step of the proposed method to calibrate the measured signal vector $\mathbf{x}^i$. Hence, the estimated parameters in the reconstruction block may be considered as the output of the DoA block in a typical radar processing. 

\subsection{Tx and Rx GPI estimation}
Exploiting the configuration of VAs, Tx and Rx imbalances can be estimated as 
\begin{subequations}\label{txrxGPI}
	\begin{align}
		\mathbf{M} &= \text{reshape}(\hat{\boldsymbol{\xi}} ^{i+1},K_{\text{r}} , K_{\text{t}}),\\
		\hat{\boldsymbol{\xi}}_{\text{t}}^{i+1} &= \frac{1}{K_{\text{r}}} \sum_{k_{\text{r}}=1}^{K_{\text{r}}} \frac{{[M[k_{\text{r}},1],...,M[k_{\text{r}},K_{\text{t}}]]^T}}{{M[k_{\text{r}},1]}},\\
		\hat{\boldsymbol{\xi}}_{\text{r}}^{i+1} &= \frac{1}{K_{\text{t}}} \sum_{k_{\text{t}}=1}^{K_{\text{t}}} \frac{{[M[1,k_{\text{t}}],...,M[K_{\text{r}},k_{\text{t}}]]^T}}{ {M[1,k_{\text{t}}]}},
	\end{align}
\end{subequations}
where $\mathbf{M}$ is an auxiliary matrix, and $\text{reshape}(\hat{\boldsymbol{\xi}} ^{i+1},K_{\text{r}} , K_{\text{t}})$, reshapes vector $\hat{\boldsymbol{\xi}} ^{i+1}$ with the size of $K_{\text{t}}K_{\text{r}} \times 1$ into matrix $\mathbf{M}$ with the size of $K_{\text{r}} \times K_{\text{t}}$. Then, Tx and Rx GPIs are estimated by computing absolute and phase values of $\hat{\boldsymbol{\xi}}_{\text{t}}^{i+1}$ and $\hat{\boldsymbol{\xi}}_{\text{r}}^{i+1}$, respectively. They are denoted as 
$\hat{\boldsymbol{\gamma}}^{i+1}_\text{t} = [\hat{\gamma}^{i+1}_\text{t}[1],\hat{\gamma}^{i+1}_\text{t}[2],...,\hat{\gamma}^{i+1}_\text{t}[K_\text{t}]]^T$, $\hat{\boldsymbol{\varphi}}^{i+1}_\text{t} = [\hat{\varphi}^{i+1}_\text{t}[1],\hat{\varphi}^{i+1}_\text{t}[2],...,\hat{\varphi}^{i+1}_\text{t}[K_\text{t}]]^T$, $\hat{\boldsymbol{\gamma}}^{i+1}_\text{r} = [\hat{\gamma}^{i+1}_\text{r}[1],\hat{\gamma}^{i+1}_\text{r}[2],...,\hat{\gamma}^{i+1}_\text{r}[K_\text{r}]]^T$, and $\hat{\boldsymbol{\varphi}}^{i+1}_\text{r} = [\hat{\varphi}^{i+1}_\text{r}[1],\hat{\varphi}^{i+1}_\text{r}[2],...,\hat{\varphi}^{i+1}_\text{r}[K_\text{r}]]^T$, respectively. 
These estimates can be used for a fault detection on Tx and Rx channels. In Section \ref{subsec:SBB}, SBB detection is introduced as one example of such fault detection tasks.

\section{Considerations} \label{considerations}
In this section, important aspects of the proposed algorithm will be explained in further details.
 
\subsection{Computational Complexity}
The computational complexity (CC) of the proposed method in each iteration can be separated into two parts corresponding to the two main blocks of the algorithm, i.e., reconstruction and adaptive filtering. Here, we consider CC of operations with floating point precision. Also a multiplication and a division are considered the same operations, as they have the same order of complexity $\mathcal{O}(1)$. 

In the reconstruction block, $K$ complex-valued multiplications are required for the predistortion step. The CC of the parameter estimation step equals to the CC of the CLEAN algorithm, mainly consisting of: a) computation of $\bar{Q}$ FFTs, where $\bar{Q}$ is an average value for the number of detected targets in the angular spectrum, and CC of the FFT is $\mathcal{O}(N\text{log}N)$, b) $\bar{Q}$ grid searches with the complexity of $\mathcal{O}(N)$ to find the dominant peak in each iteration of the CLEAN algorithm. Besides, the CC for the integration of parameters based on \eqref{s_hat} is $\bar{Q}$ times the computation of $K$ exponential functions and $K$ multiplications. 
The CC of computing an exponential function using a CORDIC algorithm is in the order of $\mathcal{O}(o_\text{t})$, where $o_\text{t}$ denotes the number of iterations of the CORDIC algorithm. Considering $o_\text{t}$ to be a fixed and small number, the CC of computing an exponential function can be simplified to $\mathcal{O}(1)$. Therefore, the CC of the reconstruction block is in the order of $\mathcal{O}(\bar{Q}N\text{log}N)$, when $N \gg K$.  

The adaptive filtering block consists of $K$ single-tap NLMS estimations. The total CC of this block equals to the computations required in \eqref{update_equation}, \eqref{grad} and \eqref{mu}, which are $K$, $2K$ and $K+1$ complex-valued multiplications, respectively. The normalization step requires $K$ complex-valued multiplications, and the detrending step requires $K$ inverse tangent computations to extract the phases, $2K$ real-valued multiplications corresponding to \eqref{ls}, $K$ real-valued multiplications for the trend recovery and removal, and finally $K$ absolute value computations, each with CC of order $\mathcal{O}(1)$. 
The CC of an inverse tangent using a CORDIC algorithm, similar to that of an exponential function, is in the order of $\mathcal{O}(1)$. Hence, the CC of the adaptive filtering block is in the order of $\mathcal{O}(K)$.

For the Tx and Rx imbalance estimation, the CC of the computations in \eqref{txrxGPI} need to be also considered. Neglecting the CC of the reshaping function, $ K_\text{t} (K_\text{r} +1) + K_\text{r}(K_\text{t}+1)$ complex-valued multiplications are required. On top of that, $K_\text{t} + K_\text{r}$ absolute value computations for gain imbalance estimation, and the same number of complex-valued multiplications and inverse tangent computations for phase imbalance estimation are required. This results in a complexity of $\mathcal{O}(K_\text{t} K_\text{r})$, or equivalently $\mathcal{O}(K)$ for Tx and Rx GPI estimation.

Comparing the CC required by different steps of the proposed algorithm, one can conclude that the CC of the reconstruction block, and more specifically the CC of the CLEAN algorithm, dominates that of other steps. Note that, due to the large zero-padding in the DoA step, the condition $N \gg K$ generally holds. Additionally, typical processors employed for the radar processing in commercial automotive radars are designed to perform FFT-based computations efficiently. Therefore, the proposed method can be effectively implemented in an automotive radar processor.

\subsection{Combining Calibration and SBB Detection} \label{subsec:SBB}
The proposed method can be incorporated to address two separate tasks of channel imbalance calibration and SBB detection. For the latter task, the Tx and Rx GPIs are continuously compared to predefined thresholds. If any of the estimates exceed their corresponding threshold, a SBB is reported. In this work, we assume the effect of a SBB on the gain imbalances is negligible, and only compare the phase imbalance estimates to a threshold denoted as $\delta$.
Although the main steps of the proposed method for both calibration and SBB detection tasks are the same, fundamental considerations need to be taken into account. These considerations rise from the requirements of each task. 
 
In the calibration task, the channel imbalances vary gradually due to factors like temperature, voltage, and aging. Due to this gradual change, the convergence rate of the calibration method is not critical. However, a high accuracy of the estimations and calibrations is required. Hence, the step size $\mu$ has to be small enough to ensure an acceptable error in the estimates. In contrast, the SBB detection task requires fast detection of phase changes caused by an SBB. In this case, the accuracy of the estimates is less critical. Therefore, the step size $\mu$ has to be large enough to ensure an acceptable SBB detection speed.  
 
\begin{figure}[t]
	\centering
	\includegraphics{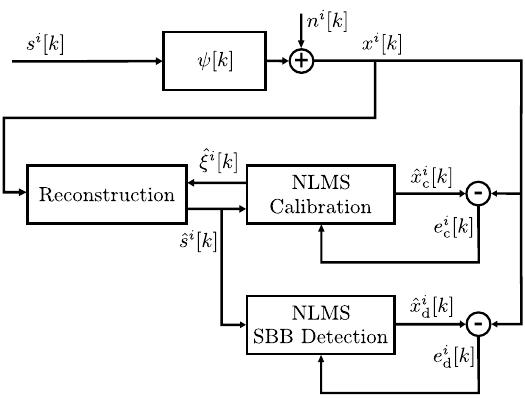}
	\caption{Combination of calibration and SBB detection tasks.}
	\label{combination}
\end{figure}

Applying different step sizes, affects the reconstruction block in the predistortion step. Therefore, one might assume that not only separate adaptive filtering blocks but also separate reconstruction blocks are required, suggesting the two tasks must be addressed separately. However, even though the adaptive filtering blocks differ due to different step sizes, a single reconstruction block can be shared between the two tasks, as shown in Fig.~\ref{combination}. The idea here is that we can apply the updated channel imbalances from the calibration task into the predistortion step within the shared reconstruction block. Although applying a lower step size results in a predistortion with lower update rate, which eventually reduces the SBB detection speed, the CC of the proposed method is reduced by almost 50\%. 

\subsection{Convergence}\label{convergence}
Considering the adaptive filtering with noisy input model in Fig.~\ref{inputnoise_bd}, a proof of convergence for the proposed method is provided. Without loss of generality, we derive the proof for only one of the channels. Inserting \eqref{grad} into \eqref{update_equation}, and then inserting \eqref{x} into the result for each $k$ gives
\begin{subequations}\label{b1}
	\begin{align}
		\hat{\psi}^{i+1} [k]  &= \hat{\psi}^i [k] - \mu^i (\hat{ {s}}^i [k])^\ast(\hat{\psi}^i [k]  \hat{ {s}}^i [k]-  {x}^i [k])\\
		\begin{split}
			&=  \hat{\psi}^i [k] - \mu^i (\hat{ {s}}^i [k])^\ast(\hat{\psi}^i [k]  \hat{ {s}}^i [k] -  {\psi [k]}   {s} ^i [k] \\
			&  \hspace{4.65cm}-  {n}^i [k]). \label {b1_1}
		\end{split}
	\end{align}
\end{subequations}
Then, \eqref{r} is used to rewrite \eqref{b1_1} as
\begin{subequations}\label{b2}
	\begin{align}
		\begin{split}
			\hat{\psi}^{i+1} [k]  &= \hat{\psi}^i [k] - \mu^i (\hat{ {s}}^i [k])^\ast (\hat{\psi}^i [k]  \hat{ {s}}^i [k]  \\
			& \hspace{1.2cm}  -{\psi [k]}   (\hat{s}^i [k] - r^i [k])-  {n}^i [k]) \label {b2_1}
		\end{split}\\
		\begin{split}
			&= \hat{\psi}^i [k] - \mu^i (\hat{ {s}}^i [k])^\ast ( (\hat{\psi}^i [k] - \psi [k])  \hat{ {s}}^i [k] \\
			& \hspace{1.2cm}  +  \psi [k] r^i [k] -  {n}^i [k]) \label {b2_2}
		\end{split}\\
		\begin{split}
			&= \hat{\psi}^i [k] - \mu^i  (\hat{\psi}^i [k] - \psi [k]) |\hat{ {s}}^i [k])|^2  \\
			& \hspace{1.2cm}  -  \mu^i (\hat{ {s}}^i [k])^\ast \psi [k] r^i [k] \\
			& \hspace{1.2cm} +  \mu^i (\hat{ {s}}^i [k])^\ast {n}^i [k]. \label {b2_3}
		\end{split}
	\end{align}
\end{subequations}

By subtracting both sides of \eqref{b2_3} by $\psi [k]$, and defining the estimation error as $\tilde{\psi} ^i[k] = \hat{\psi}^i [k] - \psi [k]$, we arrive at 
\begin{subequations}\label{b3}
	\begin{align}
		\begin{split}
			\tilde{\psi}^{i+1}[k] &= \tilde{\psi}^i[k] - \mu^i  \tilde{\psi}^i[k] |\hat{ {s}}^i [k]|^2  \\
			& \hspace{1.2cm}  -  \mu^i (\hat{ {s}}^i [k])^\ast \psi [k] r^i [k] \\
			& \hspace{1.2cm} +  \mu^i (\hat{ {s}}^i [k])^\ast {n}^i [k] \label {b3_1}
		\end{split}\\
		\begin{split}
			 &= \tilde{\psi}^i[k] (1 - \mu^i |\hat{ {s}}^i [k]|^2) \\
			& \hspace{1.2cm}  -  \mu^i (\hat{ {s}}^i [k])^\ast \psi [k] r^i [k] \\
			& \hspace{1.2cm} +  \mu^i (\hat{ {s}}^i [k])^\ast {n}^i [k]. \label {b3_2}
		\end{split}
\end{align}
\end{subequations}

Then, inserting \eqref{mu} into \eqref{b3_2}, and taking the expectation at both sides yields 
\begin{equation}\label{expectation}
		\begin{split}
			E \left[ \tilde{\psi}^{i+1}[k] \right] &= E \left[\tilde{\psi}^i[k] \left(1 - \frac{\mu_0}{(\hat{\mathbf{s}}^i)^H \hat{\mathbf{s}}^i} |\hat{ {s}}^i [k]|^2\right)\right] \\
			& - E \left[\frac{\mu_0}{(\hat{\mathbf{s}}^i)^H \hat{\mathbf{s}}^i} (\hat{ {s}}^i [k])^\ast \psi [k] r^i [k] \right] \\
			& + E \left[ \frac{\mu_0}{(\hat{\mathbf{s}}^i)^H \hat{\mathbf{s}}^i} (\hat{ {s}}^i [k])^\ast {n}^i [k] \right],
		\end{split}
\end{equation}
where $E[\cdot]$ denotes the ensemble expected value. Assuming $\tilde{\psi}^i[k]$ and $1 - \frac{\mu_0}{(\hat{\mathbf{s}}^i)^H \hat{\mathbf{s}}^i} |\hat{ {s}}^i [k]|^2$ are uncorrelated, the first term in \eqref{expectation} can be written in the form of $E \left[\tilde{\psi}^i[k] \right] \left(1 - \mu_0 E \left[ \frac{|\hat{{s}}^i [k]|^2}{(\hat{\mathbf{s}}^i)^H \hat{\mathbf{s}}^i} \right]\right)$. Also, assuming $\frac{\mu_0}{(\hat{\mathbf{s}}^i)^H \hat{\mathbf{s}}^i} (\hat{ {s}}^i [k])^\ast$ and $n^i[k]$ are statistically independent, the third term in \eqref{expectation} can be written in the form of $E \left[ \frac{\mu_0}{(\hat{\mathbf{s}}^i)^H \hat{\mathbf{s}}^i} (\hat{ {s}}^i [k])^\ast \right] E \left[{n}^i [k] \right]$, and since ${n}^i [k]$ is zero-mean, this term vanishes. Finally, assuming $\psi [k]$ is independent from other parameters, \eqref{expectation} is simplified as
\begin{equation}\label{expectation2}
	\begin{split}
		E \left[ \tilde{\psi}^{i+1}[k] \right] &= E \left[\tilde{\psi}^i[k] \right] \left(1 - \mu_0 E \left[ \frac{ |\hat{{s}}^i [k]|^2}{(\hat{\mathbf{s}}^i)^H \hat{\mathbf{s}}^i} \right]\right) \\
		& - \mu_0 E \left[\frac{(\hat{ {s}}^i [k])^\ast r^i [k]}{(\hat{\mathbf{s}}^i)^H \hat{\mathbf{s}}^i}  \right] \psi [k]. 
	\end{split}
\end{equation}

If $\mu_0$ is chosen in a way that $|1 - \mu_0 E \left[ \frac{ |\hat{{s}}^i [k]|^2}{(\hat{\mathbf{s}}^i)^H \hat{\mathbf{s}}^i} \right]| < 1$, the term $E \left[ \tilde{\psi}^{i+1}[k] \right]$ in \eqref{expectation2} converges to
\begin{equation}\label{bias}
	b[k] = b_0[k] \psi[k],
\end{equation}
where
\begin{equation}\label{bc}
	b_0[k] = \frac{1}{E \left[ \frac{ |\hat{{s}}^i [k]|^2}{(\hat{\mathbf{s}}^i)^H \hat{\mathbf{s}}^i} \right]} E \left[\frac{(\hat{s}^i[k])^\ast r^i[k]}{{(\hat{\mathbf{s}}^i)^H \hat{\mathbf{s}}^i}} \right].
\end{equation}

Finally, the bias vector is defined as $\mathbf{b} = [b[1], b[2],...,b[K]]^T$. From \eqref{bias} and \eqref{bc}, one can verify that a smaller reconstruction error $\mathbf{r}^i$, or equivalently smaller correlation between the reconstruction error $\mathbf{r}^i$ and the reconstructed signal $\hat{\mathbf{s}}^i$ results in a smaller estimation bias. However, as we will show in the results sections this consideration is addressed to a great degree by utilizing the CLEAN algorithm.  

To simplify the stability condition $|1 - \mu_0 E \left[ \frac{ |\hat{{s}}^i [k]|^2}{(\hat{\mathbf{s}}^i)^H \hat{\mathbf{s}}^i} \right]| < 1$, one can approximate $E \left[ \frac{ |\hat{{s}}^i [k]|^2}{(\hat{\mathbf{s}}^i)^H \hat{\mathbf{s}}^i} \right] \approx \frac{1}{K}$ and therefore, $|1 -  \frac{\mu_0}{K}| < 1$. Thereby, $\mu_0$ is chosen in the range of $(0, \hspace{1mm} 2K)$ to guarantee stability. However, to ensure a desired steady state estimation error, $\mu_0$ is chosen significantly smaller than the upper bound. 


\subsection{More on the bias vector} \label{bias+}
In this subsection, the bias vector $\mathbf{b}$ is investigated further. Denoting the signals corresponding to the detected and undetected targets in the parameter estimation step as $\mathbf{s}^i_1$ and $\mathbf{s}^i_2$, respectively, one can write 
\begin{equation}\label{s1s2}
	\mathbf{s}^i = \mathbf{s}^i_1 + \mathbf{s}^i_2,
\end{equation}
where ${\mathbf{s}}^i_1 = [s^i_1 [1] , s^i_1 [2], ...,s^i_1 [K]]^T$, and 
\begin{equation}\label{s1}
	s^i_1 [k] = \sum_{{q_1} = 1}^{Q_1} \alpha_{1,{q_1}}
	e^{\text{j} (2 \pi f_{1,{q_1}}^\theta  k )},
\end{equation}
where  $Q_1$ is the number of correctly detected targets, and $\alpha_{1,{q_1}}$ and $f_{1,{q_1}}^\theta$ are the amplitude and frequency of the $q_1$th correctly detected target.

Also, the reconstructed signal $\hat{\mathbf{s}}^i$ can be written in the form of 
\begin{equation}\label{s1s2hat}
	\hat{\mathbf{s}}^i = \hat{\mathbf{s}}^i_1 + \mathbf{s}^i_3,
\end{equation}
where $\hat{\mathbf{s}}^i_1$ is the signal corresponding to the correctly detected targets, and $\mathbf{s}^i_3$ is the signal corresponding to extra detected targets, which are typically referred to as ghost targets. Each element of $\hat{\mathbf{s}}^i_1 = [\hat{s}^i_1 [1] , \hat{s}^i_1 [2], ...,\hat{s}^i_1 [K]]^T$ is written as
\begin{equation}\label{s1hat}
	\hat{s}^i_1 [k] = \sum_{{q_1} = 1}^{Q_1} (\alpha_{1,{q_1}} + \Delta\alpha_{1,{q_1}}) 
	e^{\text{j} (2 \pi (f_{1,{q_1}}^\theta + \Delta f_{1,{q_1}}^\theta) k )},
\end{equation}
where $\Delta \alpha_{1,{q}}$ and $\Delta f_{q_1}^\theta$ are the amplitude and frequency estimation error. 

By inserting \eqref{s1s2} and \eqref{s1s2hat} into \eqref{r}, the term ${(\hat{s}^i[k])^\ast r^i[k]}$ can be rewritten as
\begin{subequations}\label{sr}
	\begin{align}
		(\hat{s}^i[k])^\ast &r^i[k] = (\hat{s}^i[k])^\ast (\hat{s}^i[k] - s^i[k]) \\
		\begin{split}
			&= (\hat{s}^i_1 [k] + s^i_3 [k])^\ast \\
			& \left( (\hat{s}^i_1 [k] + s^i_3 [k]) - (s^i_1 [k] + s^i_2 [k]) \right)	
		\end{split}\\
		\begin{split}
			&= (\hat{s}^i_1 [k])^\ast \left( \hat{s}^i_1 [k] + s^i_3 [k] - s^i_1 [k] - s^i_2 [k] \right)\\
			&+ (s^i_3 [k])^\ast \left( \hat{s}^i_1 [k] + s^i_3 [k] - s^i_1 [k] - s^i_2 [k] \right).
		\end{split} \label{sr_c}
	\end{align}
\end{subequations}

By combining \eqref{sr_c} with the term $E \left[\frac{(\hat{s}^i[k])^\ast r^i[k]}{{(\hat{\mathbf{s}}^i)^H \hat{\mathbf{s}}^i}} \right]$ in \eqref{bc} we have
\begin{subequations}\label{Esr}
	\begin{align}
		\begin{split}
			& E \left[\frac{(\hat{s}^i[k])^\ast r^i[k]}{{(\hat{\mathbf{s}}^i)^H \hat{\mathbf{s}}^i}} \right]\\ 
			& = E \left[ \frac{(\hat{s}^i_1 [k])^\ast \left( \hat{s}^i_1 [k] + s^i_3 [k] - s^i_1 [k] - s^i_2 [k] \right)} {{(\hat{\mathbf{s}}^i)^H \hat{\mathbf{s}}^i}} \right. \\
			& \left. + \frac{(s^i_3 [k])^\ast \left( \hat{s}^i_1 [k] + s^i_3 [k] - s^i_1 [k] - s^i_2 [k] \right)} {{(\hat{\mathbf{s}}^i)^H \hat{\mathbf{s}}^i}} \right]
		\end{split}\\
		\begin{split}
			& = E \left[ \frac{(\hat{s}^i_1 [k])^\ast \hat{s}^i_1 [k]} {{(\hat{\mathbf{s}}^i)^H \hat{\mathbf{s}}^i}} \right]
			  + E \left[ \frac{(\hat{s}^i_1 [k])^\ast  s^i_3 [k] } {{(\hat{\mathbf{s}}^i)^H \hat{\mathbf{s}}^i}} \right]\\
			& - E \left[ \frac{(\hat{s}^i_1 [k])^\ast  s^i_1 [k] } {{(\hat{\mathbf{s}}^i)^H \hat{\mathbf{s}}^i}} \right]
			  - E \left[ \frac{(\hat{s}^i_1 [k])^\ast  s^i_2 [k] } {{(\hat{\mathbf{s}}^i)^H \hat{\mathbf{s}}^i}} \right]\\
			& + E \left[ \frac{(s^i_3 [k])^\ast \hat{s}^i_1 [k]} {{(\hat{\mathbf{s}}^i)^H \hat{\mathbf{s}}^i}} \right]
			+ E \left[ \frac{(s^i_3 [k])^\ast  s^i_3 [k] } {{(\hat{\mathbf{s}}^i)^H \hat{\mathbf{s}}^i}} \right]\\
			& - E \left[ \frac{(s^i_3 [k])^\ast  s^i_1 [k] } {{(\hat{\mathbf{s}}^i)^H \hat{\mathbf{s}}^i}} \right]
			- E \left[ \frac{(s^i_3 [k])^\ast  s^i_2 [k] } {{(\hat{\mathbf{s}}^i)^H \hat{\mathbf{s}}^i}} \right].
		\end{split} \label{Esr_b}
	\end{align}
\end{subequations}

Since the signals $\mathbf{s}^i_1$, $\mathbf{s}^i_2$ and $ \mathbf{s}^i_3$ generally contain distinct and random complex amplitude and frequency components, it can be shown that their elements are stochastically orthogonal. The same statement holds for $\hat{\mathbf{s}}^i_1$, $\mathbf{s}^i_2$ and $ \mathbf{s}^i_3$. We also assume that their pairwise products are uncorrelated with $\frac{1} {{(\hat{\mathbf{s}}^i)^H \hat{\mathbf{s}}^i}}$. Hence, the corresponding correlation terms in \eqref{Esr_b} vanish, and it is simplified as

\begin{equation}\label{Esr2}
		\begin{split}
			E \left[\frac{(\hat{s}^i[k])^\ast r^i[k]}{{(\hat{\mathbf{s}}^i)^H \hat{\mathbf{s}}^i}} \right] 
			 &= E \left[ \frac{ |\hat{s}^i_1 [k]|^2} {{(\hat{\mathbf{s}}^i)^H \hat{\mathbf{s}}^i}} \right]
			 + E \left[ \frac{ |s^i_3 [k]|^2 } {{(\hat{\mathbf{s}}^i)^H \hat{\mathbf{s}}^i}} \right]\\
			&- E \left[ \frac{(\hat{s}^i_1 [k])^\ast  s^i_1 [k] } {{(\hat{\mathbf{s}}^i)^H \hat{\mathbf{s}}^i}} \right].
		\end{split}
\end{equation}

The term $(\hat{s}^i_1 [k])^\ast  s^i_1 [k] $ in \eqref{Esr2} can be expanded as

\begin{equation}\label{Ess}
		\begin{split}
			&(\hat{s}^i_1 [k])^\ast  s^i_1 [k] \\
			&= \left( \sum_{{q_1} = 1}^{Q_1} (\alpha_{1,{q_1}} + \Delta\alpha_{1,{q_1}}) 
			e^{\text{j} (2 \pi (f_{1,{q_1}}^\theta + \Delta f_{1,{q_1}}^\theta) k )} \right)^\ast \\ 
			& \hspace{0.3cm} \left( \sum_{{q_1}' = 1}^{Q_1} \alpha_{1,{q_1}'} e^{\text{j} (2 \pi f_{1,{q_1}'}^\theta  k )} \right).
		\end{split}
\end{equation}

Considering constant phases of the targets, frequencies of the targets, and the frequency estimation errors in \eqref{Ess} have symmetric and zero-mean distributions, the corresponding expected value in \eqref{Esr2} $E \left[ \frac{(\hat{s}^i_1 [k])^\ast  s^i_1 [k] } {{(\hat{\mathbf{s}}^i)^H \hat{\mathbf{s}}^i}} \right]$ will be a real value. As the other two terms in \eqref{Esr2} are real, the coefficient $b_0[k]$ is also real. 

Also, after convergence, one can write

\begin{subequations}\label{}
	\begin{align}
			E \left[ \tilde{\psi}^{i}[k] \right] & = b_0[k] \psi[k], \\
			E \left[ \hat{\psi}^i [k] \right] & =  \psi [k] + b_0[k] \psi[k] \\
			& = (1 + b_0[k]) \psi[k].
		\end{align}
\end{subequations}

Due to the fact that $1 + b_0[k]$ is a real value, one can write $\angle E \left[ \hat{\psi}^i [k] \right] = \angle \psi[k]$. And since the estimations form a cluster around their average value, one can show $\angle E \left[ \hat{\psi}^i [k] \right] \approx  E \left[ \angle \hat{\psi}^i [k] \right]$. See \cite{kim2021phase} for similar argument. Consequently, $ E \left[ \angle \hat{\psi}^i [k] \right] =  \angle \psi[k]$. This derivation specifies that the phase imbalance estimation is approximately unbiased.

\section{Simulation results}\label{results}

Consider an automotive FMCW radar with $K_\text{t} = 3$ Tx antennas and $K_\text{r}=4$ Rx antennas forming a uniform linear VA of size $K=12$ with inter-element spacing of $d=\frac{\lambda}{2}$. We carry out the simulations at MCU level and after the signal extraction block in Fig.~\ref{block_processing}. For that, 2000 signal vectors with random parameters are generated according to the signal model in \eqref{s} and \eqref{x}. In this model, two sets of targets are considered.
For each signal vector, the number of targets in the first set is selected from the values $1,2,...,5$, with probabilities of 40\%, 30\%, 15\%, 10\%, and 5\%, respectively. Similarly, the number of targets in the second set is selected from the values $0,1,...,3$, with a uniform distribution. The absolute values of the complex amplitudes for the targets in the first set are uniformly selected from [-10, 0] dB. For the targets in the second set, these values are chosen from [-20 -10) dB below the dominant peak in the first set. The goal of considering the second set is to model the smaller targets that are challenging to detect in both single and multiple target cases. Additionally, The constant phase values are selected uniformly from [$-\pi$  $\pi$]. And, the directions of the targets are selected uniformly from [$-90^o$  $90^o$]. For the parameter estimation in the reconstruction block, we employ a CLEAN algorithm, incorporating $N=1024$ points FFTs and a termination criterion threshold of $P = -15$ dB. It should be noted that, throughout this section, we initially detrend the injected phase imbalances. The effects of a linear trend has been discussed in Section \ref{subsec:Imbalance}.

\begin{figure}[]
	\centering
	\includegraphics[scale=0.45]{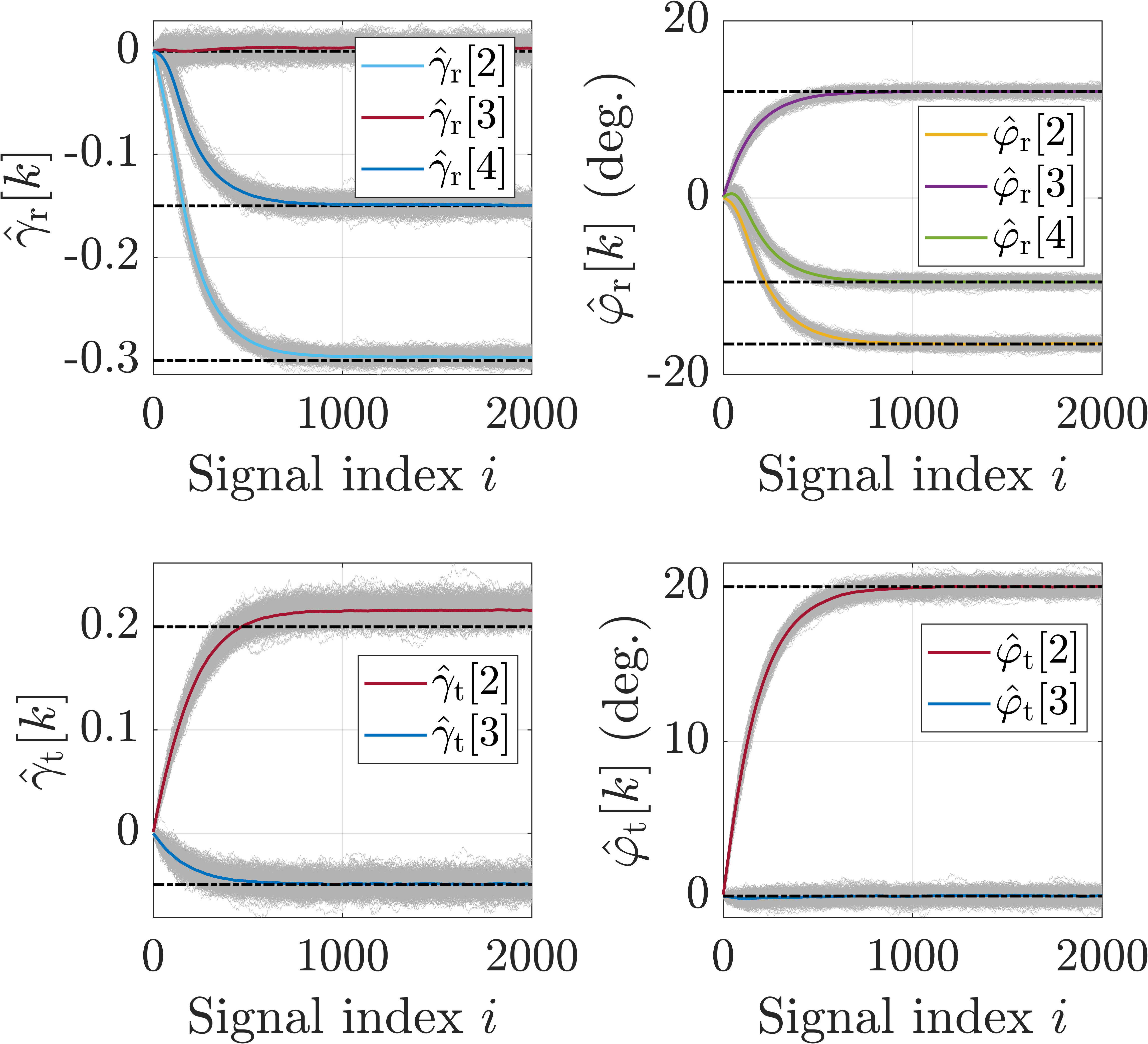}
	\caption{Tx and Rx GPI estimates. }
	\label{sim1txrx}
\end{figure}

In the first simulation, the SNR in \eqref{x} is set to 20 dB. Also, the GPIs are considered constant across all iterations, whereby one iteration refers to the processing of one signal vector. For this simulation, we set the NLMS step size to $\mu_0 = 0.1$. Fig.~\ref{sim1txrx} illustrates the Tx and Rx GPI estimation results for 1000 MCSs. The dashed black lines denote the injected imbalances, and the gray curves show all MCSs. Also, the colored curves show the average values of all MCSs for each channel. As can be seen, the algorithm converges after approximately 1000 iterations. To gain a practical understanding of the time duration of these 1000 iterations, we consider an automotive radar operating with a frame rate of 20 frames per second. Assuming each radar frame contains 10 signal vectors, each corresponding to a detected peak in the R-D map, 1000 iterations correspond to 5 seconds.

\begin{figure}[]
	\centering
	\includegraphics[scale=0.45]{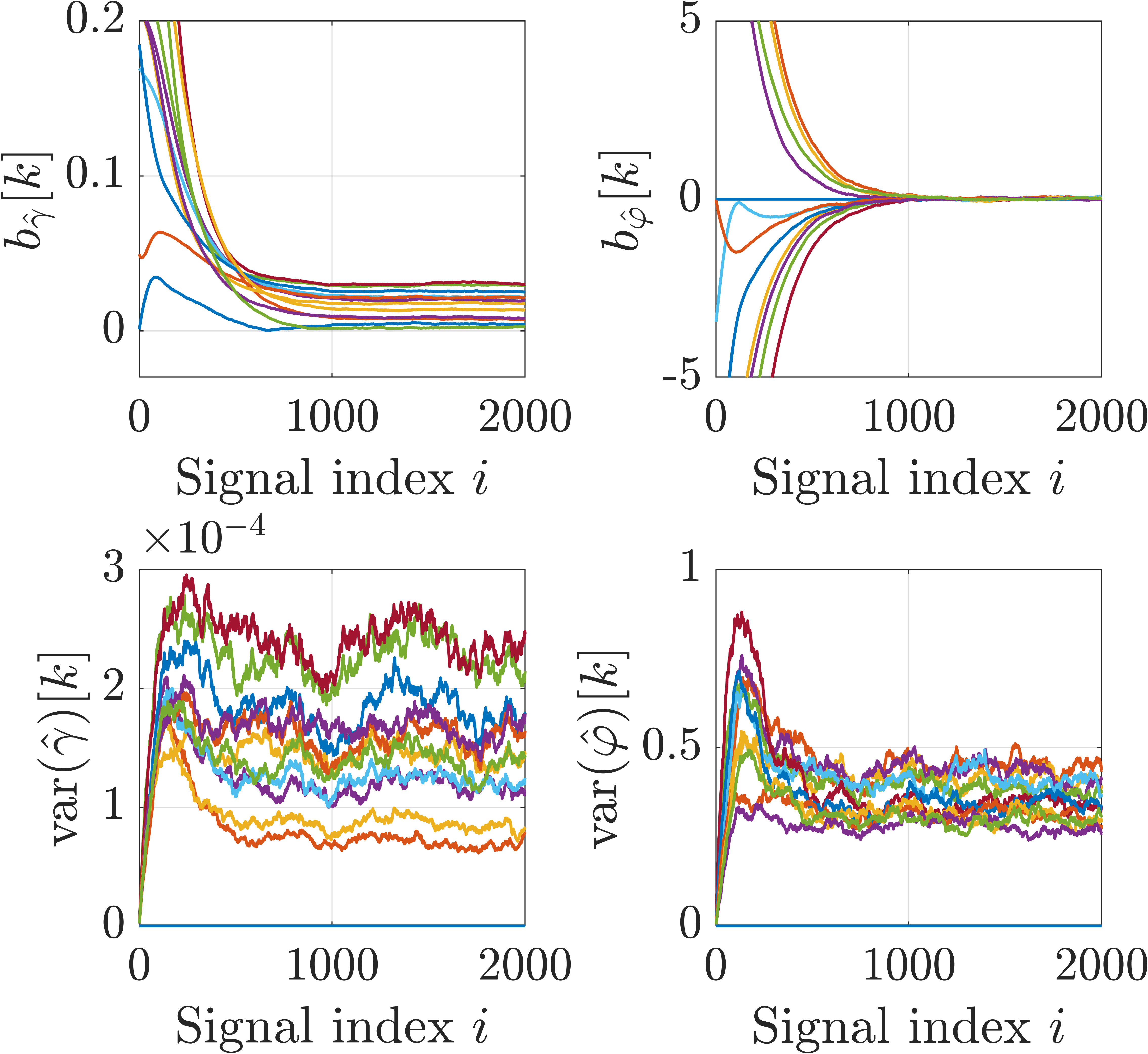}
	\caption{Bias (in degrees) and variance behavior of the GPI estimates for all VA channels.}
	\label{sim1biasvar}
\end{figure}

In Fig.~\ref{sim1biasvar}, the convergence behavior of the proposed method for all $K=12$ VA channels is investigated. In the upper subfigures, the biases in the GPI estimates denoted as $b_{\hat{\gamma}}[k]$ and $b_{\hat{\varphi}}[k]$ are illustrated. Clearly, the bias in the phase estimates tends to zero, as it was shown in Section \ref{bias+}, and the bias in the gain estimates is negligible. In the lower subfigures, the empirical variances of GPI estimates over all MCSs denoted as $\text{var}(\hat{\gamma})[k]$ and $\text{var}(\hat{\varphi})[k]$ are depicted. These variance plots illustrate the variation of the estimates around the converged value for each VA channel. This can also be seen from the variation of gray plots in Fig.~\ref{sim1txrx}.

\begin{figure}[t]
	\centering
	\includegraphics[scale=0.45]{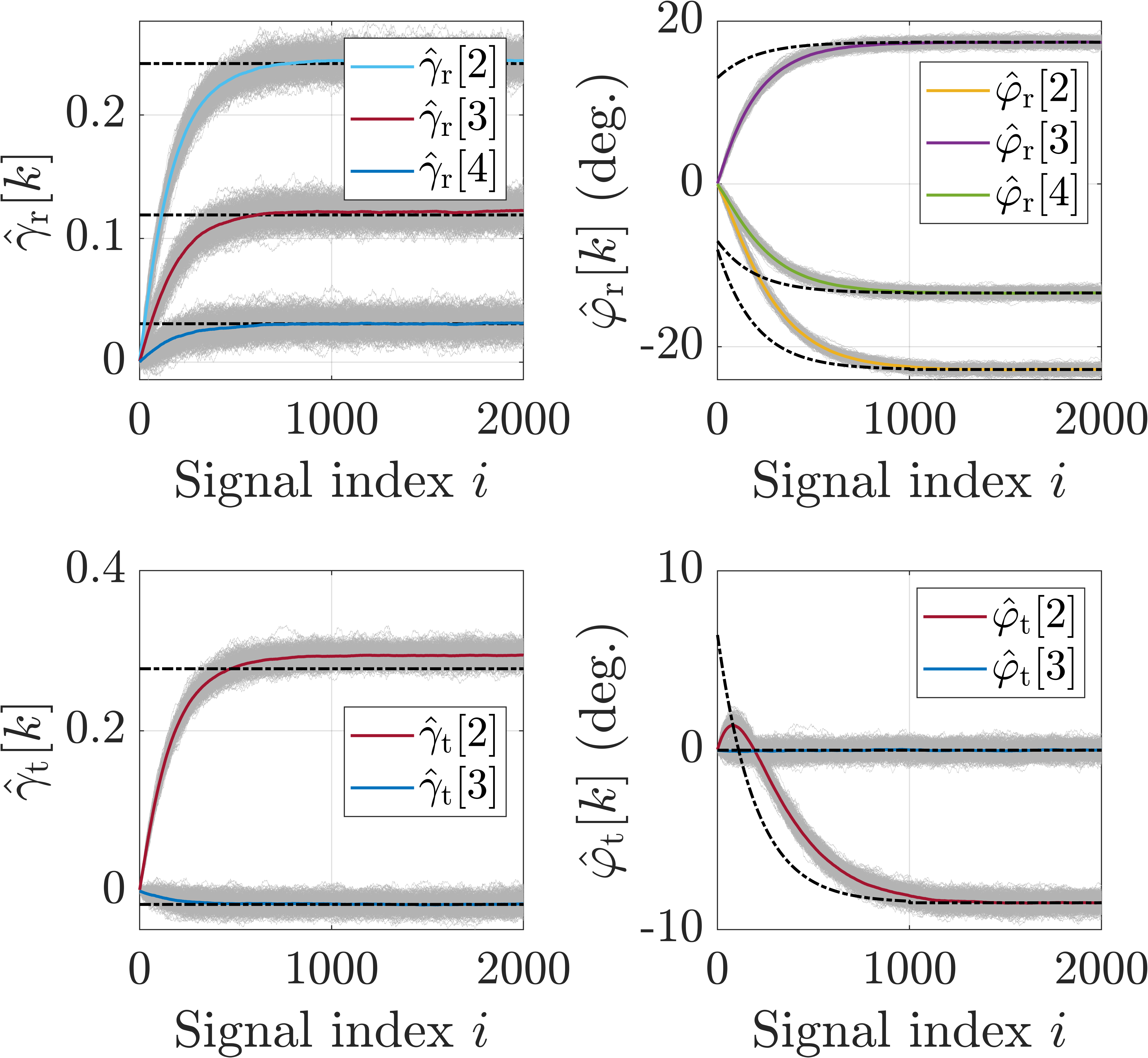}
	\caption{Tx and Rx GPI estimates during the heat-up process of the radar.}
	\label{sim2txrx}
\end{figure}

One critical scenario for imbalance estimation occurs right upon activation of the radar, when the internal temperature of the radar, and more specifically the MMIC, significantly increases in a short period of time. Consequently, the imbalances may change accordingly. To simulate this behavior, we consider the set up in the first simulation, except that we change the phase imbalance model as follows. For the first 1000 iterations, equivalent to 5 seconds with the aforementioned assumptions, exponential functions are used to model the phase imbalances during the heat-up process of the radar. For the second 1000 iterations, constant values equal to the final values of the first half are considered. Also, the gain imbalances are considered constant across all iterations. Fig.~\ref{sim2txrx} depicts the results of this simulation. Although the algorithm follows the imbalances closely after roughly 1000 iterations, a faster tracking of imbalances might be desired in the first iterations of this scenario. 

\begin{figure}[t]
	\centering
	\includegraphics[scale=0.45]{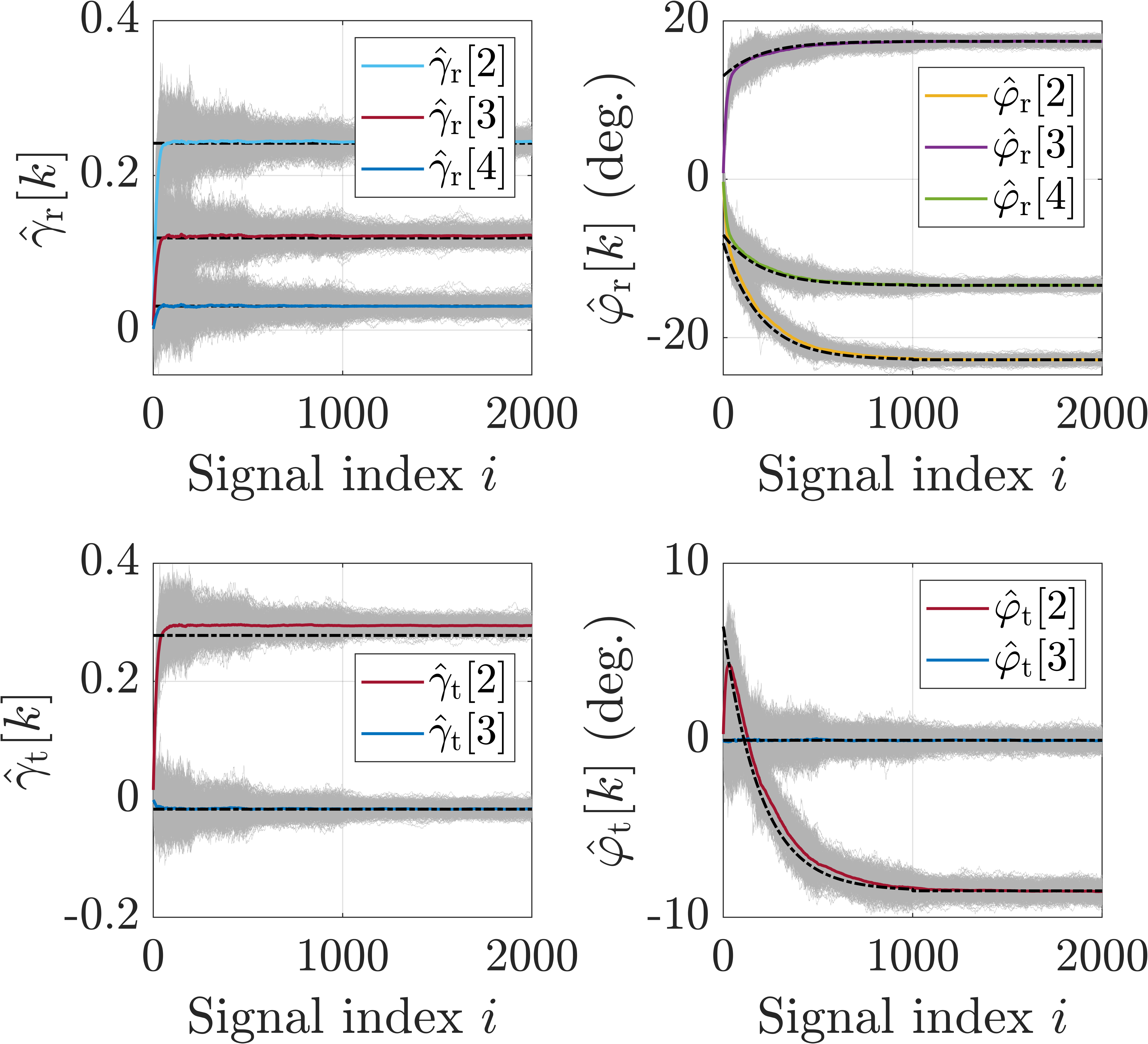}
	\caption{Tx and Rx GPI estimates during the heat-up process of the radar, when multiple normalized step sizes are used for faster convergence in the initial iterations.}
	\label{sim2variablemu}
\end{figure}

One approach to achieve a faster convergence is to modify the step size $\mu_0$ over iterations. For instance, one may use a larger step size in the heat-up process. Here, we modify the step size in multiple stages as follows. From iteration 1 to 50, the step size is set to $\mu_0 = 1$, from iteration 51 to 200, $\mu_0 = 0.8$, from iteration 201 to 500, $\mu_0 = 0.4$, from iteration 501 to 1000, $\mu_0 = 0.2$, and from iteration 1001 onward, $\mu_0 = 0.1$. The estimation results are illustrated in Fig.~\ref{sim2variablemu}, from which one can conclude that the algorithm can follow the imbalances closely after 50 iterations, which is 20 times faster than the implementation without varying the normalized step size.

\begin{figure}[t]
	\centering
	\includegraphics[scale=0.43]{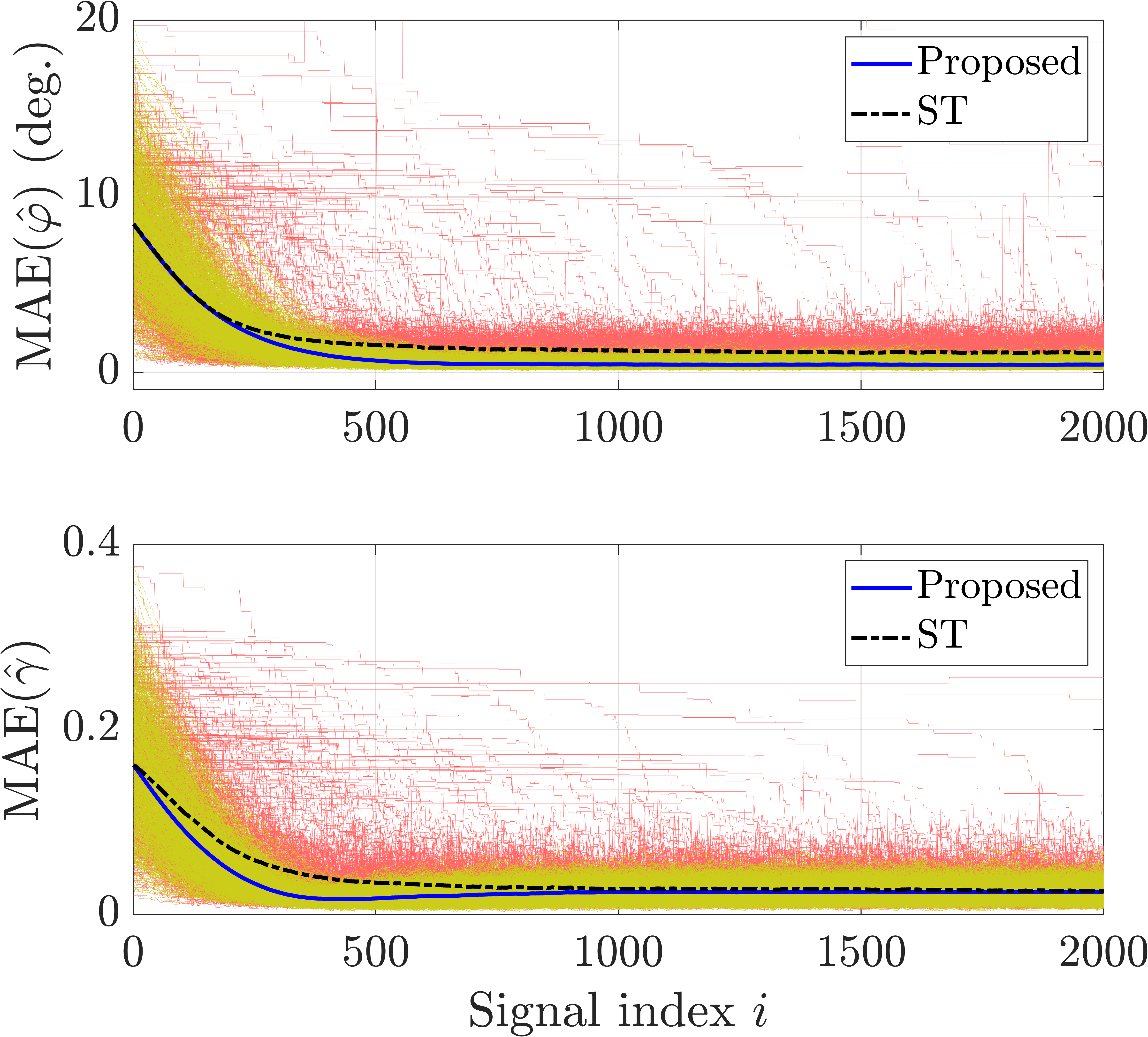}
	\caption{ Comparison of GPI estimation performance between the proposed and the ST methods.}
	\label{sim3comp}
\end{figure}

In the next simulation, the GPI estimation as well as the calibration performance of the proposed method are compared to those of the method in \cite{af}, which utilizes only single targets. Hereafter, this method is referred to as ST. The injected GPIs are considered constant over the iterations, and random over 1000 MCSs in the range of ${\varphi_\text{t}, \varphi_\text{r}} \in [-20^o \hspace{1mm} 20^o]$ and ${\gamma_\text{t}, \gamma_\text{r}} \in [-0.2 \hspace{1mm} 0.2]$. Fig.~\ref{sim3comp} illustrates the empirical mean absolute phase error $\text{MAE}(\hat{\varphi})$, in the upper plot, and the empirical mean absolute gain error $\text{MAE}(\hat{\gamma})$, in the lower plot, between the injected GPIs and the estimated ones over all VA channels and over all MCSs. In this figure, the green and red curves represent the individual MCSs results for the proposed and the ST methods, respectively, where the mean is computed only over the VA channels. The blue and the black curves indicate the mean values over all MCSs for the proposed and the ST methods, respectively. From this figure, one can observe that the proposed method leads to a lower $\text{MAE}(\hat{\varphi})$ and $\text{MAE}(\hat{\gamma})$ compared to those of the ST method. Additionally, the ST method's GPI estimation error in multiple MCSs is significantly high, resulting in a substantially poor calibration performance. The fundamental reason for this poor behavior is the ST method's limitation to the scenarios that contain enough number of single targets. Hence, in the MCSs that lack enough number of single targets, the ST method cannot converge.

\begin{figure}[t]
	\centering
	\includegraphics[scale=0.45]{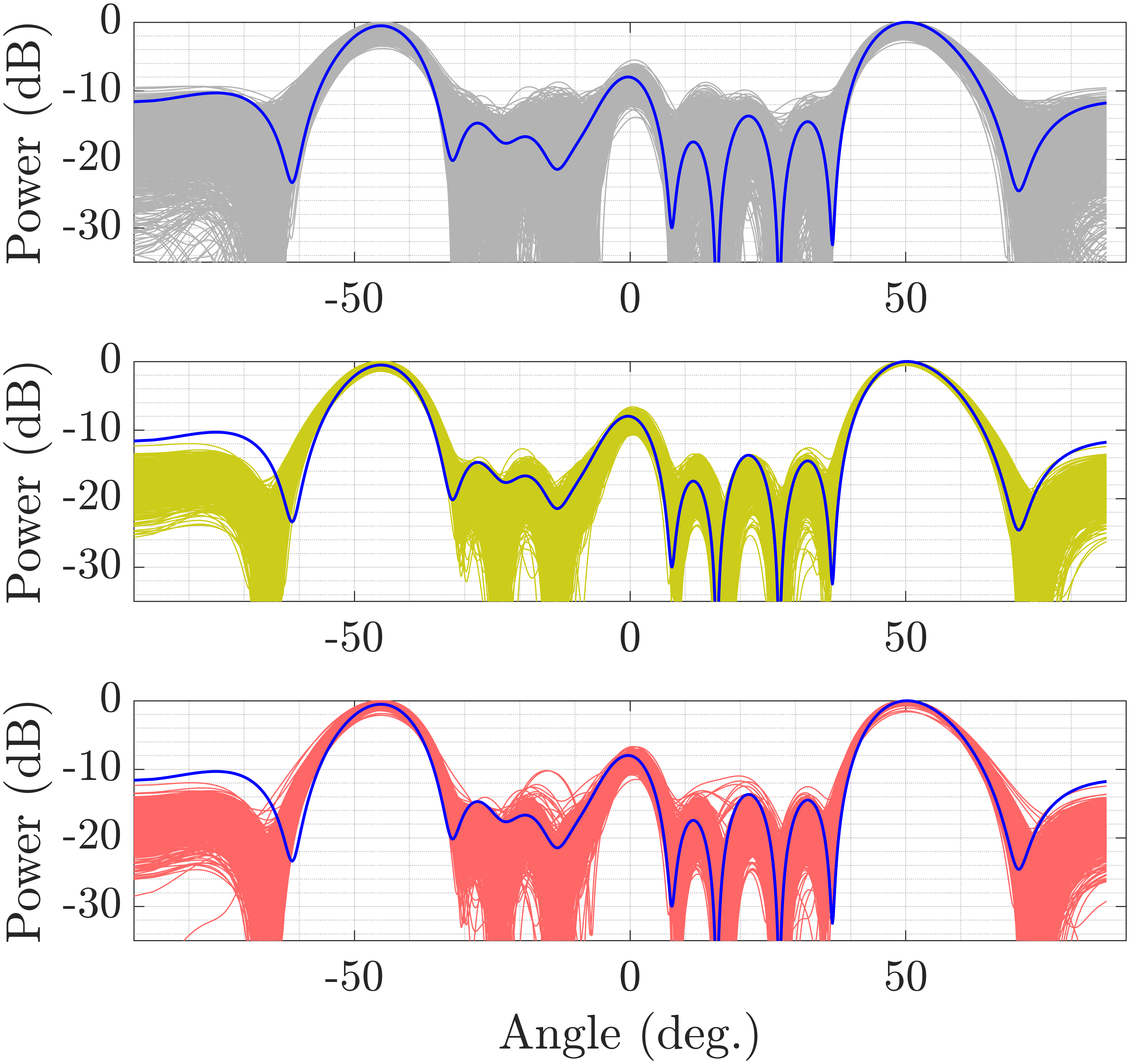}
	\caption{Comparison of calibration performance in terms of SLLS between the proposed (middle plot) and ST (lower plot) methods. The upper plot depicts the uncalibrated angular spectra, and the blue curves in all plots represent the ideal angular spectrum. }
	\label{sim3spec}
\end{figure}

After the estimation of the imbalances in each MCS, i.e., after iteration 2000, the estimates are exploited to calibrate a signal vector containing three targets as considered in \ref{subsec:Imbalance} for the simulation in Fig.~\ref{imbspec}.
Fig.~\ref{sim3spec} illustrates the calibration performance of the proposed and the ST methods. In this figure, the spectra of the signal vector before calibration (upper plot), after calibration by the proposed method (middle plot) and after calibration by the ST method (lower plot), over all MCSs are illustrated. The blue curve in all three plots represents the ideal spectrum corresponding to the imbalance-free signal vector. As can be seen, the proposed method suppresses the maximum sidelobes, from roughly -9 dB to around -13 dB, i.e., 4 dB SLL suppression (SLLS). On the other hand, although the ST method shows a roughly similar maximum SLLS, a poor performance in SLLS in a number of MCSs can be observed. This behavior of the ST method in some MCSs was noted even with drastically smaller step sizes.

\begin{figure*}[t]
	\centering
	\includegraphics[scale=0.5]{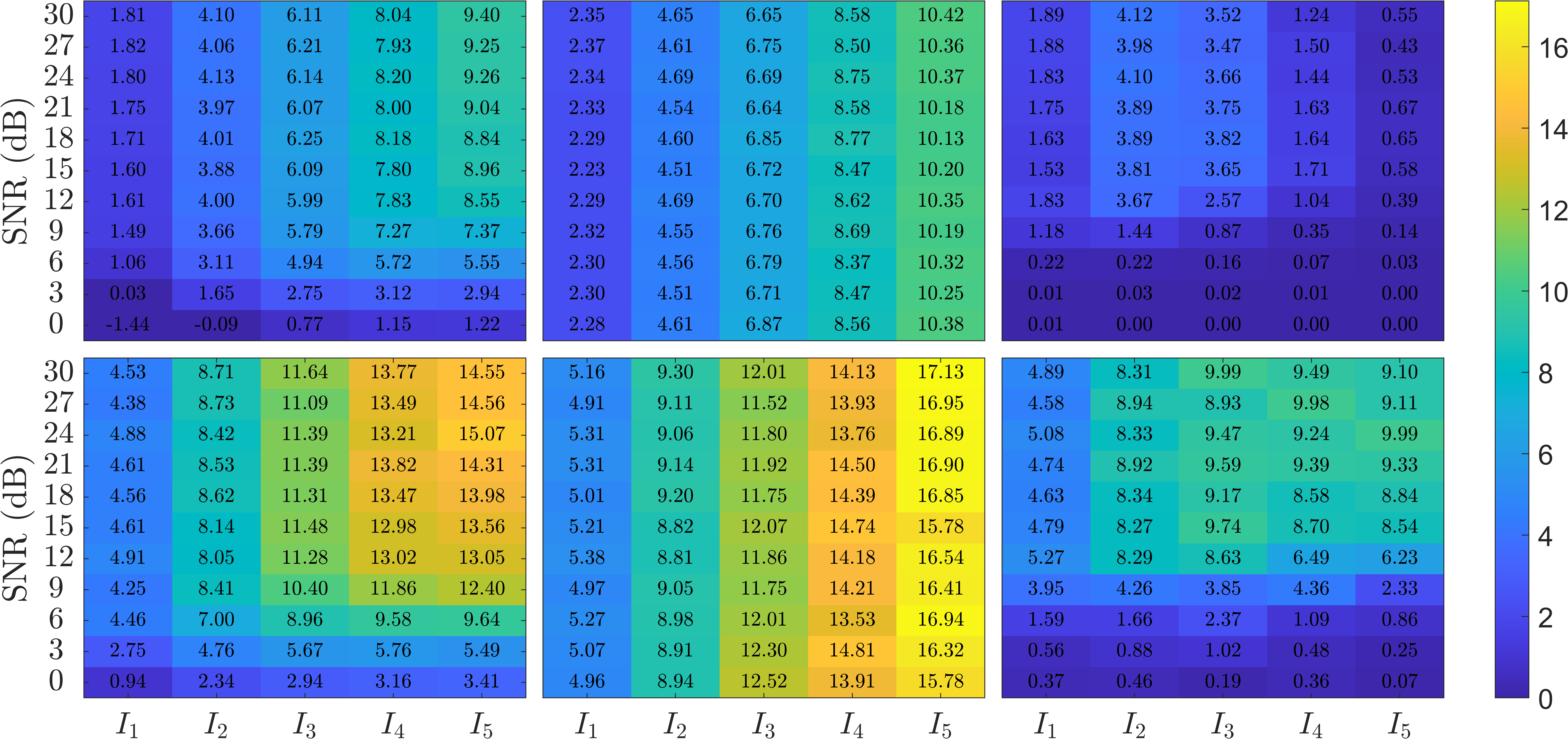}
	\caption{Comparison of the SLLS of the proposed method (left heat maps) and the ST method (right heat maps) with the ideal calibration (middle heat maps) over multiple SNR and imbalance levels. The upper and lower heat maps represent the empirical mean and maximum SLLS, respectively. }
	\label{snr_imb}
\end{figure*}

To investigate the influence of SNR and level of imbalances on the SLLS performance of the proposed method, a new simulation is devised. In this simulation, 5 levels of injected imbalances denoted as $I_l$, where $l=1,2,...,5$, are considered. For each $I_l$, GPIs are generated randomly in the range of ${\varphi_\text{t}, \varphi_\text{r}} \in [-l\times10^o \hspace{2mm} l\times10^o]$ and ${\gamma_\text{t}, \gamma_\text{r}} \in [-l\times0.1 \hspace{2mm} l\times0.1]$ over 1000 MCSs. After the estimation of the imbalances, the estimates are employed to calibrate a signal vector containing a single target at $-20^o$, and distorted by the corresponding imbalances. For comparison purposes, we generate this signal vector noise-free. 
Fig.~\ref{snr_imb} compares the SLLS of the proposed method (left heat maps) and the ST method (right heat maps) with that of the ideal calibration (middle heat maps), where the injected imbalances are used for calibration. In this figure, upper and lower heat maps represent the empirical mean and maximum SLLSs over MCSs, respectively.
The following conclusions can be derived from this simulation. 
The proposed method demonstrates excellent SLLS performance in both mean and maximum SLLS cases across all imbalance levels and SNRs higher than 6 dB.
In contrast, the ST method exhibits a weak performance in low SNR area, specifically up to 12 dB, in both mean and maximum SLLS, and also for imbalance levels equal to or greater than $I_3$ in mean SLLS. The reason is that due to the low SNRs and/or large imbalance levels, where the ST method struggles to identify sufficient single targets and therefore fails to estimate the imbalances accurately. 
Such a superior performance of the proposed method in various ranges of SNRs and imbalances indicates the reliability of the method in different scenarios, which is essential for applications with high-level FuSa requirements.

\begin{figure}[t]
	\centering
	\includegraphics[scale=0.5]{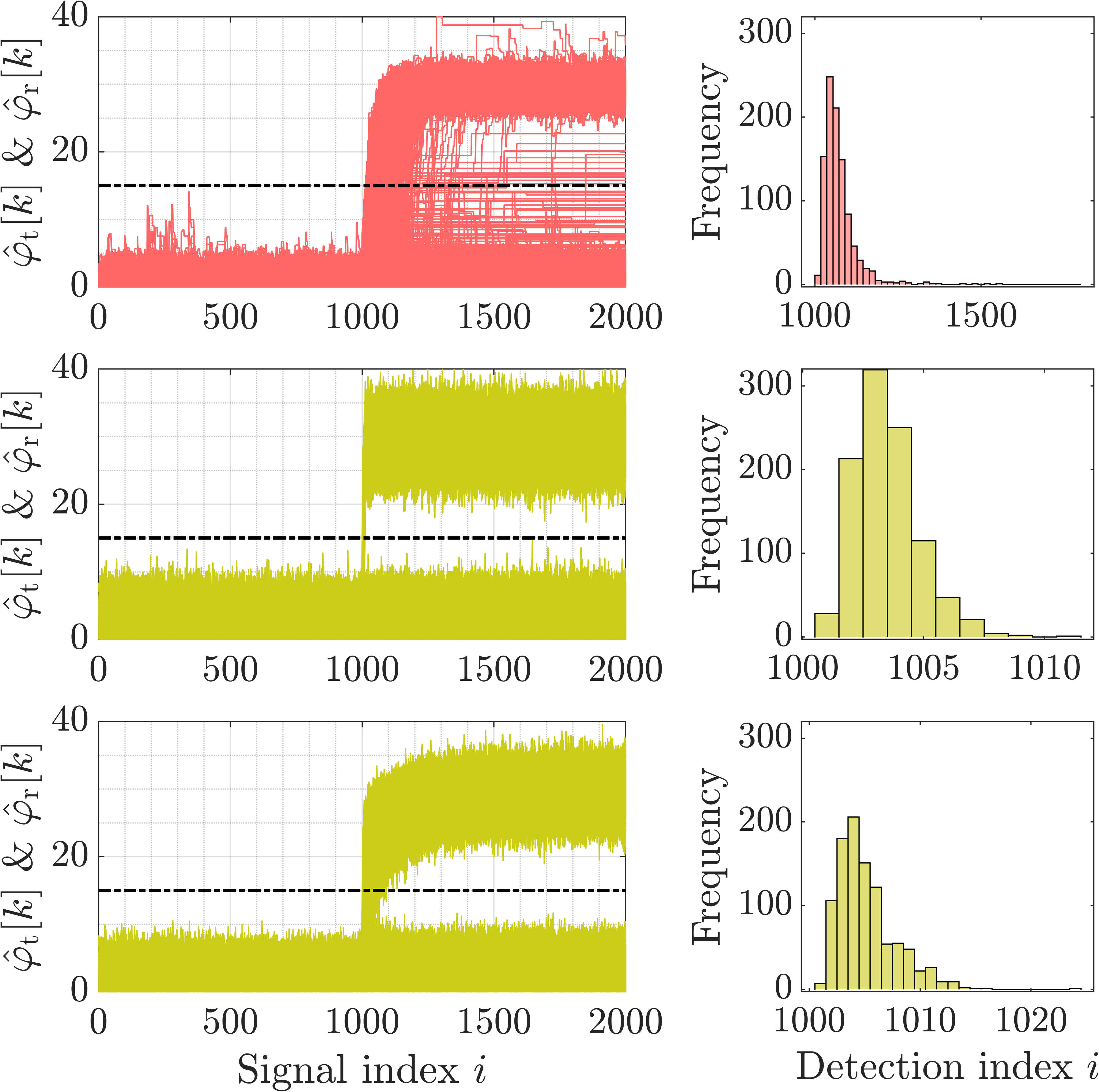}
	\caption{ Comparison of SBB detection performance between the ST method (upper plots), the proposed method when only the SBB structure is considered (middle plots), and the combined structure (lower plots). The unit of the values on the vertical axis in the left-hand side plots is degrees.}
	\label{sbbd}
\end{figure}

In the next simulation, the SBB detection performance of the proposed method is examined. To this end, a $30^o$ phase offset on the third Rx channel is considered as the SBB model \cite{bb_EKG}. It is also assumed that the SBB occurs at iteration 1000.
To detect the SBB, we increase the constant step size to $\mu_0 = 3$ in order to track the phase change faster compared to that in the calibration task. Also, we set the SBB detection threshold as half of the expected SBB-induced phase offset, i.e., $\delta=15^o$. 
The left hand side plots in Fig.~\ref{sbbd} depict the results of the Tx and Rx phase imbalance estimation over 1000 MCSs. The upper plot shows the performance of the ST method, the middle plot depicts the proposed method applied solely for the SBB detection task, and the lower plot represents the SBB detection in the combined structure. For the combined structure, the step sizes for the calibration and SBB detection tasks are similar to those in the previous simulations. In these plots, the black dashed line represents the threshold. In the right hand side plots, the histogram plots of the SBB detection indices, i.e., the index at which the phase crosses the threshold for the first time, are depicted correspondingly. Note that the limits of the detection index axis are not the same.
According to these plots, the ST method requires an average of 71 iterations and a maximum of more than 500 iterations to detect the SBB. While, the proposed method, when applied solely to the SBB detection task, requires an average of 5 iterations, and a maximum of 12 iterations to detect the SBB. In the combined structure, these values are slightly higher, with average and maximum of 7 and 25 iterations, respectively. 
One can note, that reducing the computational complexity by utilizing the combined structure comes with the price of slightly slower SBB detection. Considering the assumptions of 10 targets per radar frame, and 20 frames per second, in the worst case scenario, these two structures require 2 and 3 radar frames, or equivalently 100 and 150 milliseconds, respectively, for a SBB detection.

It is also worth mentioning that in all provided simulations, targets in different radar frames were generated independently. Conversely, in practice, the scenery evolves slowly relative to the conventional radar frame rates. This characteristic provides inherent dependency and traceability between corresponding targets across different radar frames. Therefore, in scenarios lacking single targets, the ST method remains blind for multiple consecutive radar frames, while our method maintains continuous channel imbalance estimation. This further distinguishes our method from the ST method under real-world conditions. Furthermore, both the proposed and ST methods can leverage tracking algorithms for more accurate estimation of targets' parameters.

\section{Measurement results}\label{measurement_results}
In this section the performance of the proposed method in real-world conditions is investigated. To this end, a measurement using a 77 GHz automotive radar with a virtual array of size 12 and $d=0.6\lambda$ was carried out in an indoor parking garage with moving objects in the scene. This measurement includes 176 radar frames, and a threshold of 20 dB SNR is used to detect the peaks in the R-D map of each frame. The signal vectors at each detected peak are extracted and input to the proposed and the ST methods for GPI estimation. Since the ground-truth values of the GPIs are not available, two consequent estimations are conducted. In the first estimation, the GPIs of the measurements are estimated. Afterwards, over 1000 MCSs, random GPIs in the range of ${\varphi_\text{t}, \varphi_\text{r}} \in [-20^o \hspace{1mm} 20^o]$ and ${\gamma_\text{t}, \gamma_\text{r}} \in [-0.2 \hspace{1mm} 0.2]$ are added to the extracted signal vectors. Consequently, the second imbalance estimation is conducted. Then, the resulting estimates are divided by the first estimates to derive an estimate of the artificial GPIs. In order to decorrelate the first and second estimates, signal vectors at the odd and even iterations are separated and input to the first and second estimations, respectively. It should be noted that although no EoL calibration is carried out before the estimations, the effect of the mutual coupling is vanished as a result of the relative estimation of the artificial GPIs. 

\begin{figure}[t]
	\centering
	\includegraphics[scale=0.43]{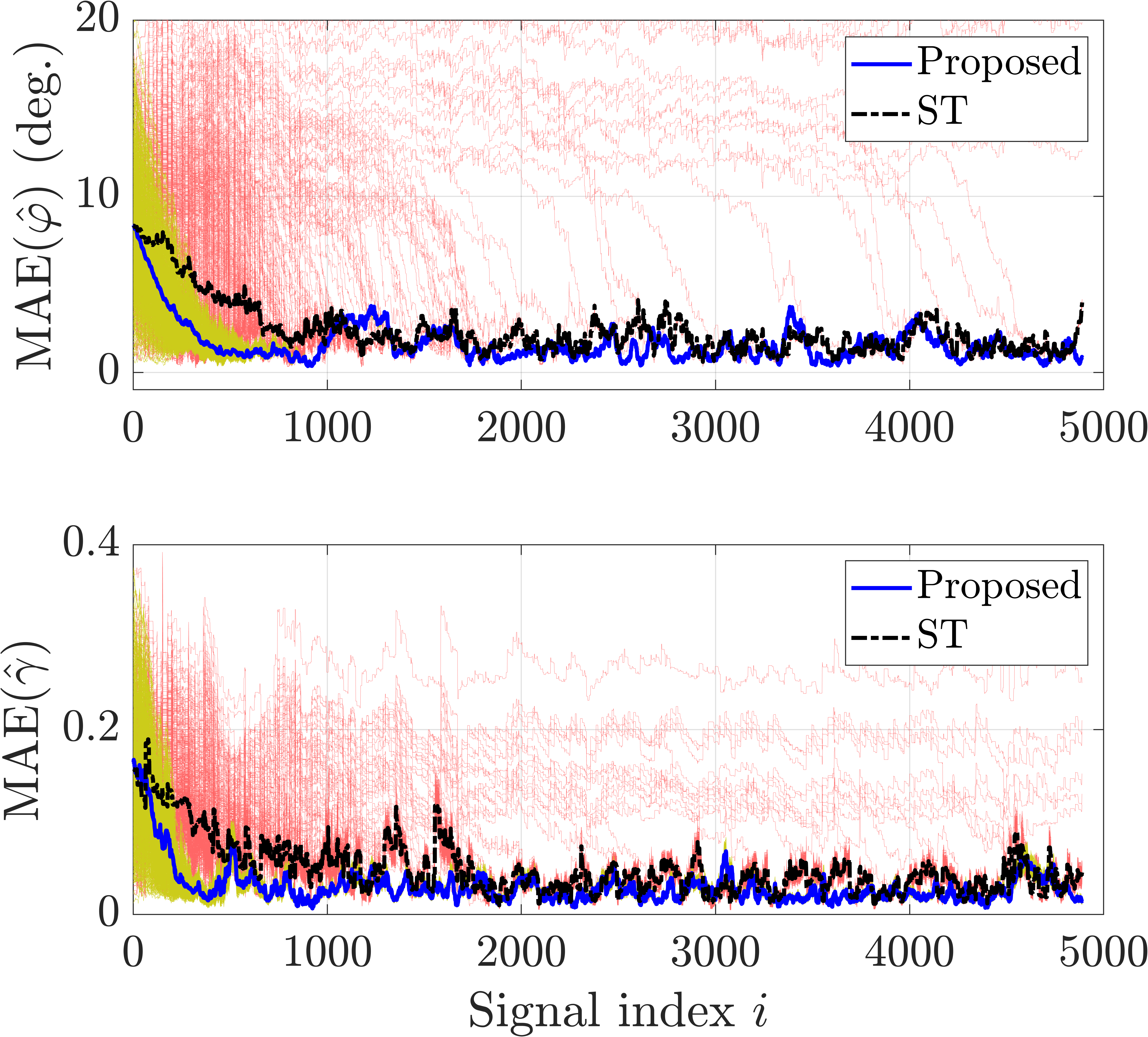}
	\caption{ Comparison of GPI estimation performance between the proposed and the ST methods using the measurement data.}
	\label{measurement_er_comp}
\end{figure}

\begin{figure}[t]
	\centering
	\includegraphics[scale=0.45]{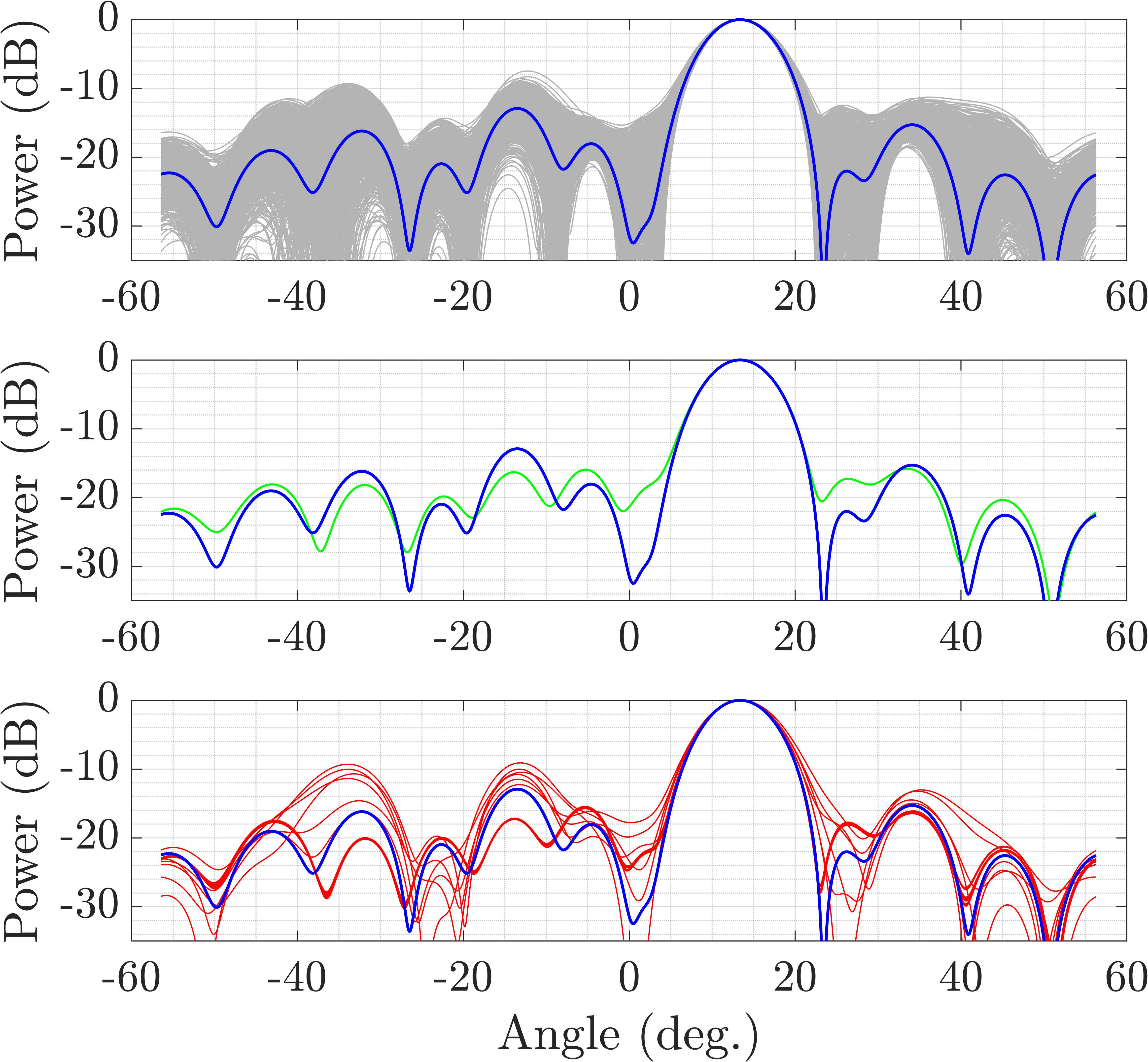}
	\caption{Comparison of calibration performance in terms of SLLS between the proposed and ST methods using the measurement data. The blue curves in all plots represent the ideal spectrum. The upper, middle and lower plots depict the uncalibrated, proposed methods' and ST methods' spectra, respectively.}
	\label{measurement_spec_comp}
\end{figure}

Fig.~\ref{measurement_er_comp} compares the $\text{MAE}(\hat{\varphi})$ and $\text{MAE}(\hat{\gamma})$ of the artificial GPIs and their estimates using the proposed and the ST method, similar to the simulation in Fig.~\ref{sim3comp}. As can be seen, similar conclusions can be made compared to the simulation results. Since these estimation errors cannot fully characterize the performance of the proposed method, for instance the bias in the estimates is removed due to the relative estimation, a calibration evaluation, similar to the simulation corresponding to Fig.~\ref{sim3spec} is carried out. For this purpose, the very last estimates of the second GPI estimation using the proposed and the ST methods are used to calibrate a signal vector deteriorated by the random imbalances in Fig.~\ref{measurement_er_comp}. Fig.~\ref{measurement_spec_comp} illustrates the results of this simulation, where the angular spectra of the deteriorated signal over all MCSs are plotted in the upper plot, the angular spectra of the calibrated signal using the proposed method in the middle plot and that using the ST method in the lower plot. The blue curve in all plots depicts the angular spectrum of the original signal vector before deterioration. From this figure, one can note a superb calibration performance of the proposed method, whereas the ST method fails to calibrate the imbalances in some MCSs. Note that all angular spectra estimated by the proposed method are the same.

\begin{figure}[t]
	\centering
	\includegraphics[scale=0.5]{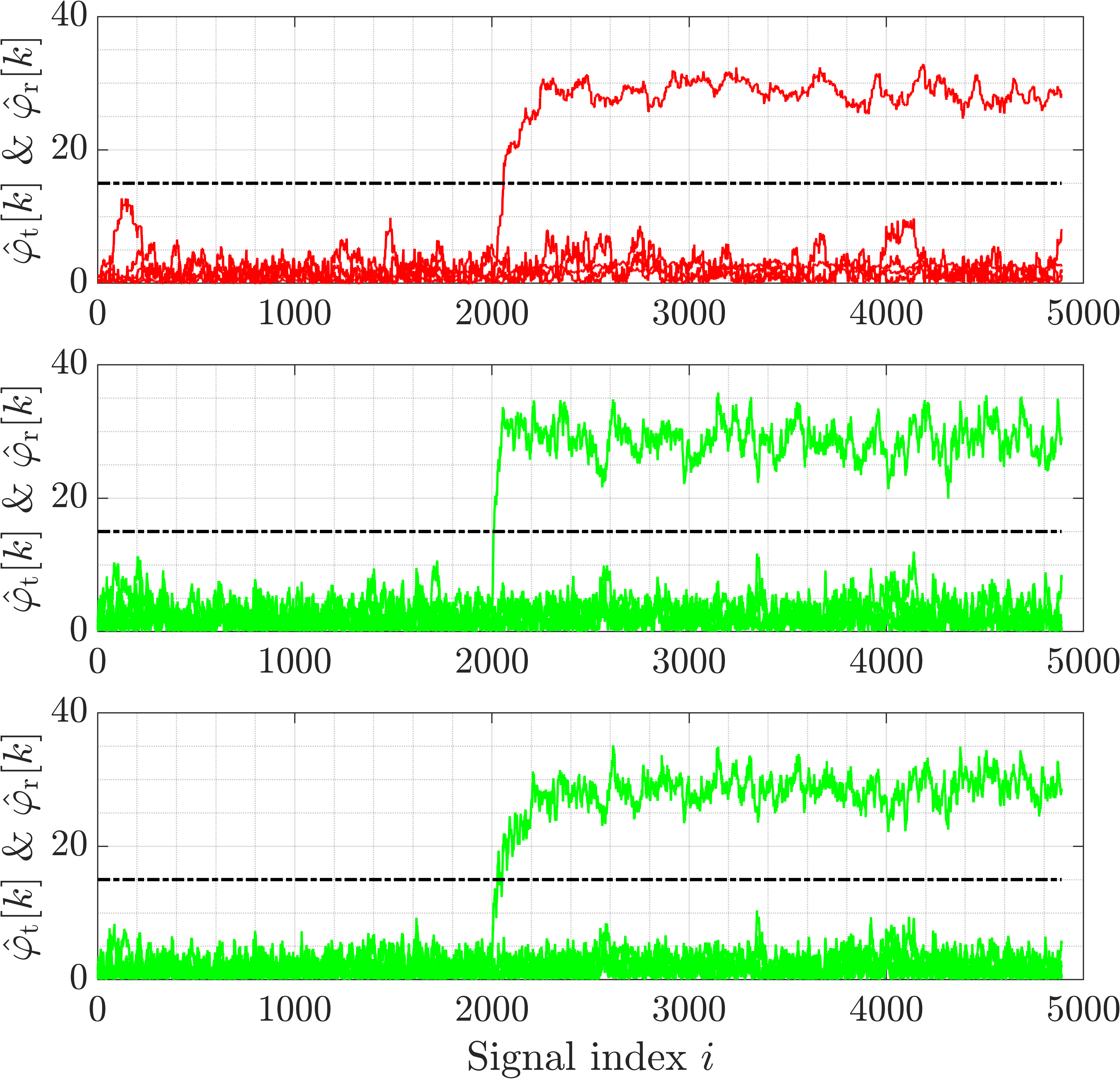}
	\caption{ Comparison of SBB detection performance between the ST method (upper plot), the proposed method when only the SBB structure is considered (middle plot), and the combined structure (lower plot) using the measurement data. The unit of the values on the vertical axis is degrees.}
	\label{sbbd_msrmnt}
\end{figure}

A similar two-step approach is used to evaluate the SBB detection performance of the proposed method using the measurement data. Here, similar to the simulation in Fig.~\ref{sbbd}, a $30^o$ phase offset on the third Rx channel is considered as the SBB model. Fig.~\ref{sbbd_msrmnt} depicts the results of the Tx and Rx artificial phase imbalance estimation. According to that, the proposed method, when applied solely to the SBB detection task and in the combined structure, requires only 9 and 28 iterations, respectively, to detect the SBB. In contrast, the ST method requires 59 iterations.

\section{Conclusion}\label{conclusion}
In this work, a method was proposed to estimate automotive radar channel imbalances online. This method exploits a cyclic approach in two steps to jointly estimate the targets' parameters as well as the channel imbalances. In the first step, a CLEAN algorithm is utilized to estimate the parameters, and in the second step, an NLMS method is exploited to estimate the channel imbalances. At each iteration of the algorithm, the estimated channel imbalances are applied to the extracted signal vectors from the R-D map to calibrate the measurement for parameter estimation. In addition, the estimates can also be used for a fault detection problem, e.g., a SBB detection, either as a separate task in parallel to calibration or in a combined structure with calibration to reduce the computational complexity. The proposed method does not require any prior knowledge of the location of targets, as the targets' parameters are estimated in the reconstruction block of the method. It is also not restricted to single targets in each R-D bin, and exploits all detected peaks on the R-D map. This makes the algorithm faster and more robust, compared to the counterpart method, specially for the SBB detection task which requires fast tracking of phase imbalances. Furthermore, simulation and measurement results showed that the proposed method outperforms the ST method in terms of SLLS and SBB detection speed. These properties, along with the low computational complexity of the NLMS method, make the proposed method a strong candidate for the online automotive radar channel imbalance estimation. 

\section*{Acknowledgments}
This work has been funded by the Linz Center of Mechatronics (LCM) GmbH as part of a K2 project. K2 projects are financed using funding from the Austrian COMET K2 programme. The COMET K2 projects at LCM are supported by the Austrian federal government, the federal state of Upper Austria, the Johannes Kepler University and all of the scientific partners which form part of the COMET K2 consortium.

\bibliographystyle{IEEEtran}
\bibliography{refs}

\end{document}